\documentclass[a4paper,notitlepage,aps,twocolumn,pra]{revtex4-1}

\usepackage[a4paper,top=2.5cm,bottom=2cm,left=2.5cm,right=2cm]{geometry}	% page margins

\usepackage{amsmath,amssymb,amsfonts} % standard AMS packages

\usepackage{bm}	% bold symbols in math mode \bm{...}
\renewcommand{\mathbf}{\bm}
\usepackage{dsfont}	% proper mathbb format
\renewcommand{\mathbb}{\mathds}	% redefine \mathbb

\usepackage{mathrsfs} % use \mathscr{} for script letters in math

\usepackage{mathtools} % for proper typesetting of := and =:

\usepackage{graphicx,float}
\usepackage[colorlinks,
	linkcolor=red,
	citecolor=blue,
	urlcolor=red]{hyperref}

\usepackage{wrapfig}	% bunch of stuff to wrap text around images
\newcommand{\fref}[1]{Fig.~\ref{#1}}
\renewcommand{\eqref}[1]{Eq.~\ref{#1}}
\newcommand{\secref}[1]{Sec.~\ref{#1}}

\renewcommand{\t}[1]{\mathrm{#1}}

\begin{document}
%\title{Closing the gap: vertical integration of a SiN nanobeam and a SiO$_2$ microdisk\\for Heisenberg-limited evanescent displacement sensing}
\title{Near-field integration of a SiN nanobeam and a SiO$_2$ microcavity\\for Heisenberg-limited displacement sensing}

\author{R. Schilling, H. Sch{\"u}tz, A. Ghadimi, V. Sudhir, D. J. Wilson, and T. J. Kippenberg}
%\author{Fellowship of the string}

\affiliation{Institute of Condensed Matter Physics, \'{E}cole Polytechnique F\'{e}d\'{e}rale Lausanne, CH-1015 Lausanne, Switzerland}
%\affiliation{\'{E}cole Polytechnique F\'{e}d\'{e}rale Lausanne (EPFL), CH-1015 Lausanne, Switzerland}

\date{\today}

\begin{abstract}
	
Placing a nanomechanical object in the evanescent near-field of a high-$Q$ optical microcavity gives access to strong gradient forces and quantum-noise-limited displacement readout, offering an attractive platform for precision sensing technology and basic quantum optics research. Robustly implementing this platform is challenging, however, as it requires separating optically smooth surfaces by $\lesssim\lambda/10$. Here we describe a fully-integrated evanescent opto-nanomechanical transducer based on a high-stress Si$_3$N$_4$ nanobeam monolithically suspended above a SiO$_2$ microdisk cavity. Employing a novel vertical integration technique based on planarized sacrificial layers, we achieve beam-disk gaps as little as 25 nm while maintaining mechanical $Q\times f>10^{12}$ Hz and intrinsic optical $Q\sim10^7$. The combined low loss, small gap, and parallel-plane geometry result in exceptionally efficient transduction, characterizing by radio-frequency flexural modes with vacuum optomechanical coupling rates of 100 kHz, single-photon cooperativities in excess of unity, and zero-point frequency (displacement) noise amplitudes of 10 kHz (fm)/$\surd$Hz. In conjunction with the high power handling capacity of SiO$_2$ and low extraneous substrate noise, the transducer operates particularly well as a sensor. Deploying it in a 4 K cryostat, we recently demonstrated a displacement imprecision 40 dB below that at the standard quantum limit (SQL) with an imprecision-back-action product $<5\cdot\hbar$. In this report we provide a comprehensive description of device design, fabrication, and characterization, with an emphasis on extending Heisenberg-limited readout to room temperature. Towards this end, we describe a room temperature experiment in which a displacement imprecision 30 dB below that at the SQL and an imprecision-back-action product $<75\cdot\hbar$ is achieved. Our results impact the outlook for measurement-based quantum control of nanomechanical oscillators and offer perspective on the engineering of functionally-integrated (``hybrid") optomechanical systems.

\end{abstract}

\maketitle

%%%%%%%%%%%%%%%%%%%%%%%%%%%%%%%%%%% BEGIN MATTER %%%%%%%%%%%%%%%%%%%%%%%%%%%%%%%%%%%%

\tableofcontents

\small{

\section{Introduction}\label{sec:introduction}

\begin{figure}[br!]
	\vspace{-5pt}
	\centering
	\includegraphics[width=.99\linewidth]{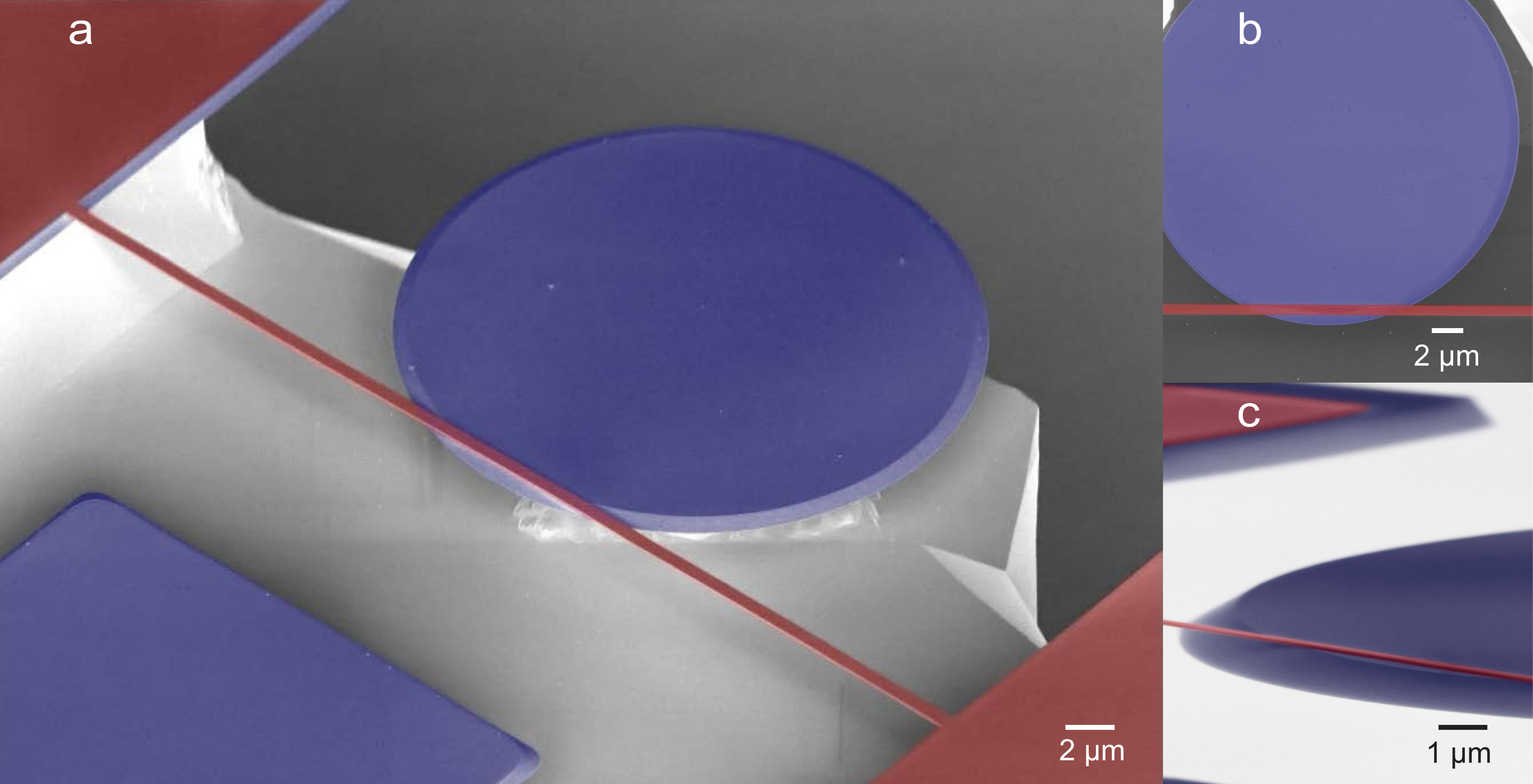}
	\caption{\label{fig:device_intro} False-colored scanning electron micrograph of the device: a high-stress Si$_3$N$_4$ (red) nanomechanical beam integrated into the evanescent mode volume of a SiO$_2$ (blue) microdisk.  Disk and beam are integrated on a Si (gray) microchip.  Subpanel b (c) highlights the lateral (vertical) positioning of the beam.}
	\vspace{-10pt}
\end{figure}
  
Nanomechanical oscillators \cite{ekinci2005nanoelectromechanical} are sensitive to weak forces and exhibit large zero-point fluctuations, making them an attractive platform for both precision sensing technology \cite{cleland1998nanometre,mamin2001sub,jensen2008atomic} and basic quantum science \cite{schwab2005putting}.  Much effort has been devoted to the development of nanomechanical transducers in the electrical domain, including single-electron transistors \cite{lahaye_approaching_2004}, atomic point contacts \cite{flowers2007intrinsic}, and superconducting microwave cavities \cite{regal_measuring_2008}. Though very successful in recent years \cite{teufel_circuit_2011}, these techniques are practically limited by the need for cryogenic operation. A comparatively powerful approach is to parametrically couple a nanomechanical oscillator to an optical cavity.  The field of a laser-driven cavity can be quantum-noise-limited at room temperature, and as such represents a practically ideal form of mechanical transducer, with read out enabled by standard interferometric techniques and actuation provided by radiation pressure.  Moreover, the finite build-up time of the cavity field allows it to do work on the mechanical element, enabling low-noise optical cooling and amplification \cite{kippenberg2008cavity}.  Investigation of these effects has led to two paradigmatic goals in the contemporary field of cavity optomechanics \cite{aspelmeyer_cavity_2015}: cooling of a solid-state mechanical oscillator to its ground state and, concomitantly, read-out of its zero-point motion with the minimal disturbance allowed by the Heisenberg uncertainty principle (due to radiation pressure shot noise (RPSN) \cite{caves_quantum-mechanical_1980}). The first challenge has been met by  several cryogenic optomechanical \cite{chan_laser_2011,verhagen2012quantum} and electromechanical systems \cite{teufel_sideband_2011} (via resolved-sideband cooling \cite{wilson2007theory}).  The latter, corresponding to a measurement at the standard quantum limit (SQL) \cite{braginsky1995quantum}, remains outstanding; however, readout noise far below the zero-point displacement has been reported \cite{wilson2015measurement,krause2015optical}, as well as RPSN dominating the thermal force \cite{purdy_observation_2013,teufel2016overwhelming}. Reaching the SQL ultimately requires a `Heisenberg-limited' displacement sensor for which the product of the read out noise and the total force noise is the minimum allowed by the uncertainty principle.  This regime has been approached to within an order of magnitude by several cryogenic systems \cite{teufel_sideband_2011,wilson2015measurement}; it also forms the basis for measurement-based quantum feedback protocols such as ground-state cooling \cite{wilson2015measurement,courty_quantum_2001} and squeezing \cite{szorkovszky2011mechanical} of an oscillator.

Efficient cavity optomechanical transduction involves co-localization of optical and mechanical modes with high $Q/\t{(mode\,volume)}$ and high optical power handling capacity.  Moreover, it is desirable that the cavity support a mechanism for efficient input/output coupling.  A diverse zoo (\fref{fig:C0_survey}) of micro- and nanoscale cavity optomechanical systems (COMS) has risen to meet these challenges, ranging from cantilevers \cite{kleckner_sub-kelvin_2006} and membranes \cite{thompson_strong_2008} coupled to Fabry-P\'{e}rot cavities to mechanically-compliant whispering-gallery-mode (WGM) microcavities \cite{schliesser_high-sensitivity_2008} and photonic crystals \cite{eichenfield_optomechanical_2009}. They generally employ two types of radiation pressure force coupling: traditional scattering-type coupling, in which the cavity field exchanges energy with the mechanical element via momentum transfer, and gradient force coupling \cite{van_thourhout_optomechanical_2010}, in which energy is exchanged via induced-dipole coupling to a field gradient.  The net effect is a parametric coupling $G=\partial\omega_\t{c}/\partial x$ between the cavity resonance frequency $\omega_\t{c}$ and the mechanical degree of freedom $x$, which expresses the force applied per intracavity photon, $\hbar\cdot G$ \cite{aspelmeyer_cavity_2015}. 

A particularly promising platform for optomechanical transduction involves placing a (dielectric) mechanical substrate next to the surface of a WGM microcavity, so that it samples its evanescent field. Since the evanescent decay length is $\sim\lambda/10$, this topology offers the opportunity for strong gradient force coupling to \emph{nanoscale} mechanical devices. It also has the virtue of naturally accommodating optical and mechanical substrates of dissimilar material and geometry, enabling separate optimization of $Q/\t{(mode\,volume)}$. Moreover, WGMs can be input/output coupled with high ideality using tapered optical fibers \cite{spillane_ideality_2003}, making them well-suited to interferometric displacement sensing. Recent work has focused on coupling of nano-beams \cite{anetsberger_near-field_2009}, -cantilevers \cite{doolin2014multidimensional}, and -membranes \cite{anetsberger_near-field_2009,cole2015evanescent} to the evanescence of WGM micro-toroids \cite{anetsberger_near-field_2009}, -spheres \cite{neuhaus2012versatile,cole2015evanescent}, and -disks \cite{gavartin_hybrid_2012,doolin2014multidimensional}, with mechanical materials ranging from (ultra low loss) high-stress Si$_3$N$_4$ \cite{anetsberger_near-field_2009} to (ultra low mass) single-layer graphene \cite{cole2015evanescent}, typically using SiO$_2$ as the optical material. Gradient force coupling as high as $G\sim2\pi\cdot100$ MHz/nm has been achieved \cite{anetsberger_near-field_2009}.  Combined with the high power handling capacity of SiO$_2$ and low extraneous displacement noise (typically thermo-refractive noise (TRN) in the cavity substrate \cite{anetsberger_measuring_2010}), optimized systems have achieved room temperature displacement imprecisions as low as $10^{-16}\,\t{m}/\sqrt{\t{Hz}}$, sufficient to in principle resolve the zero-point motion in one report \cite{anetsberger_measuring_2010}. 

Despite these advances, the full potential of evanescent cavity optomechanics has been inhibited by the difficulty of positioning the nanomechanical element within $\lambda/10\sim 100$ nm of the cavity substrate. Early systems made use of nanopositioning stages and suffered from vibrational stability \cite{anetsberger_near-field_2009}.  Gavartin \emph{et. al.} \cite{gavartin_hybrid_2012} addressed this challenge by integrating a Si$_3$N$_4$ nanobeam and a SiO$_2$ microdisk on a chip; however, due to fabrication constraints, the beam-disk separation was limited to 250 nm and the optical $Q$ was reduced by a factor of $10$.  

In this work, we discuss a novel method to monolithically integrate a high-stress Si$_3$N$_4$ thin film resonator and a SiO$_2$ microdisk cavity \emph{within the evanescent near-field}, without deteriorating the intrinsic $Q$ of either element.  The critical ingredient is a chemical mechanical polishing technique that allows integration of optically flat surfaces with sub-100 nm spacing, separated by a sacrificial film.  This procedure is used to carefully isolate Si$_3$N$_4$ and SiO$_2$ layers during wafer processing, allowing high-yield and deterministic fabrication of devices in which a nanobeam is monolithically suspended as little as 25 nm above a SiO$_2$ microdisk --- $\sim3\times$ smaller than the evanescent decay length of its WGMs --- while maintaining mechanical and optical mode qualities in excess of $10^5$ and $10^6$, respectively.  The process is compatible with e-beam lithography, thus we are able to locally pattern the beam with sub-10 nm imprecision (opening the door to stress engineering \cite{norte2015mechanical}) and laterally position it with sub-100 nm imprecision across a full 4" Si wafer.  

\begin{figure}[br!]
	\vspace{-10pt}
	\centering
	\includegraphics[width=1\linewidth]{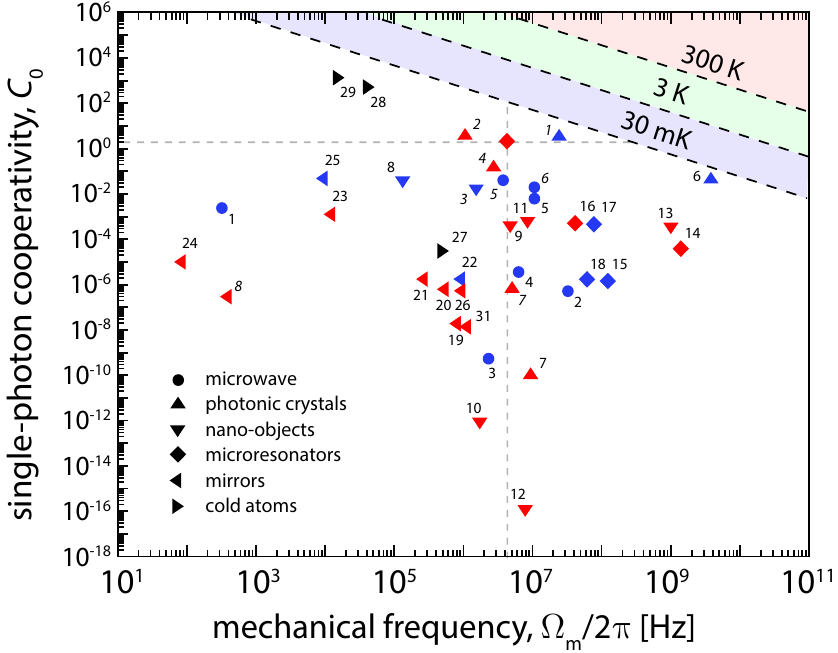}
	\caption{\label{fig:fab_flow} Survey of single-photon cooperativity $\mathcal{C}_0=4g_0^2/\kappa\Gamma_\t{m}$ for various cavity optomechanical systems \cite{safavi-naeini_squeezed_2013,krause2015optical,purdy_strong_2013,leijssen2015strong,wollman2015quantum,palomaki2013coherent,wu2014dissipative,matsumoto2015mg}, adapted with permission from \cite{aspelmeyer_cavity_2015}.  Non-italicized references are cited in \cite{aspelmeyer_cavity_2015}.  Blue and red points correspond to cryogenic (typically $T<10$ K) and room temperature experiments, respectively.  Diagonal lines indicate the condition for $\mathcal{C}_0 = n_\t{th}\approx  k_\t{B}T/\hbar\Omega_\t{m}$, for various $T$.  The reported result is highlighted with crosshairs.}\label{fig:C0_survey}
	\vspace{-5pt}
\end{figure}

A typical device is shown in \fref{fig:device_intro}, corresponding to a $60\times1\times0.06\,\mu\t{m}^3$ beam positioned 25 nm above a 0.65-$\mu$m-thick, 30-$\mu$m-diameter microdisk. By carefully varying the dimensions of the beam, the disk, and their lateral offset with respect to this nominal geometry, we achieve optomechanical coupling rates ($G$) in excess of $2\pi\cdot 1$ GHz/nm while maintaining cavity decay rates ($\kappa$) as low as $2\pi\cdot100$ MHz and radio frequency ($\Omega_\t{m}=2\pi\cdot(1-10)$ MHz) flexural beam modes with damping rates ($\Gamma_\t{m}$) as low as $2\pi\cdot 10$ Hz.  In conjunction with the small mass ($m\sim 10$ pg) and large zero-point displacement ($x_\t{zp}\equiv \sqrt{\hbar/2m\Omega_\t{m}}\sim 10$ fm) of the beam modes, the combined low-loss, small gap and parallel-plane geometry result in a vacuum optomechanical coupling rates ($g_0\equiv G\cdot x_\t{zp}$) as high as $2\pi\cdot 100$ kHz and room temperature single-photon cooperativities as high as $\mathcal{C}_0\equiv 4g_0^2/\kappa\Gamma_\t{m}=2$.  The latter is notably a factor of $10^5$ times larger than in \cite{gavartin_hybrid_2012} and on par with the state-of-the-art for both room temperature and cryogenic COMS (\fref{fig:C0_survey}).  

In conjunction with high $\mathcal{C}_0$, several features of the system make it well-suited for quantum-limited operation.  First, SiO$_2$ microcavities with the reported dimensions and internal loss readily support intracavity photon numbers of $n_\t{c}\sim 10^6$.  This enables quantum cooperativities ($C_0 n_\t{c}/n_\t{th}$) approaching unity --- a basic requirement for performing a Heisenberg-limited displacement measurement --- for a room temperature thermal occupation of $n_\t{th}\approx k_\t{B}T/\hbar\Omega_\t{m}\sim 10^6$, corresponding to $\Omega_\t{m}\sim2\pi\cdot5$ MHz. Another striking feature is the exceptionally large magnitude of the cavity frequency noise produced by zero-point motion of the mechanical oscillator, $S_\omega^{zp}(\Omega_\t{m})\equiv 4g_0^2/\Gamma_\t{m}\sim 10\,\t{kHz}/\sqrt{\t{Hz}}$.  This magnitude is many orders of magnitude larger than typical extraneous sources of noise due to laser frequency fluctuations or TRN \cite{anetsberger_measuring_2010}.  Taking advantage of these strengths, recent deployment of the device in a 4 K Helium cryostat enabled interferometric measurements with a read-out noise 43 dB below $S_\omega^\t{zp}(\Omega_\t{m})$ (corresponding to an imprecision 40 dB below that necessary to reach the SQL) and with an imprecision-back-action product of $5\cdot\hbar$, allowing active feedback cooling to near the motional ground state \cite{wilson2015measurement}.  Below, we demonstrate a measurement with an imprecision 30 dB below that at the SQL and an imprecision-back-action product of $75\cdot\hbar$, using a moderate input power of 10 $\mu$W.  Remarkably, the imprecision due to microdisk TRN \cite{anetsberger_measuring_2010} can be $20$ dB lower.

\begin{figure}[bl!]
	\vspace{-5pt}
	\centering
	\includegraphics[width=1\linewidth]{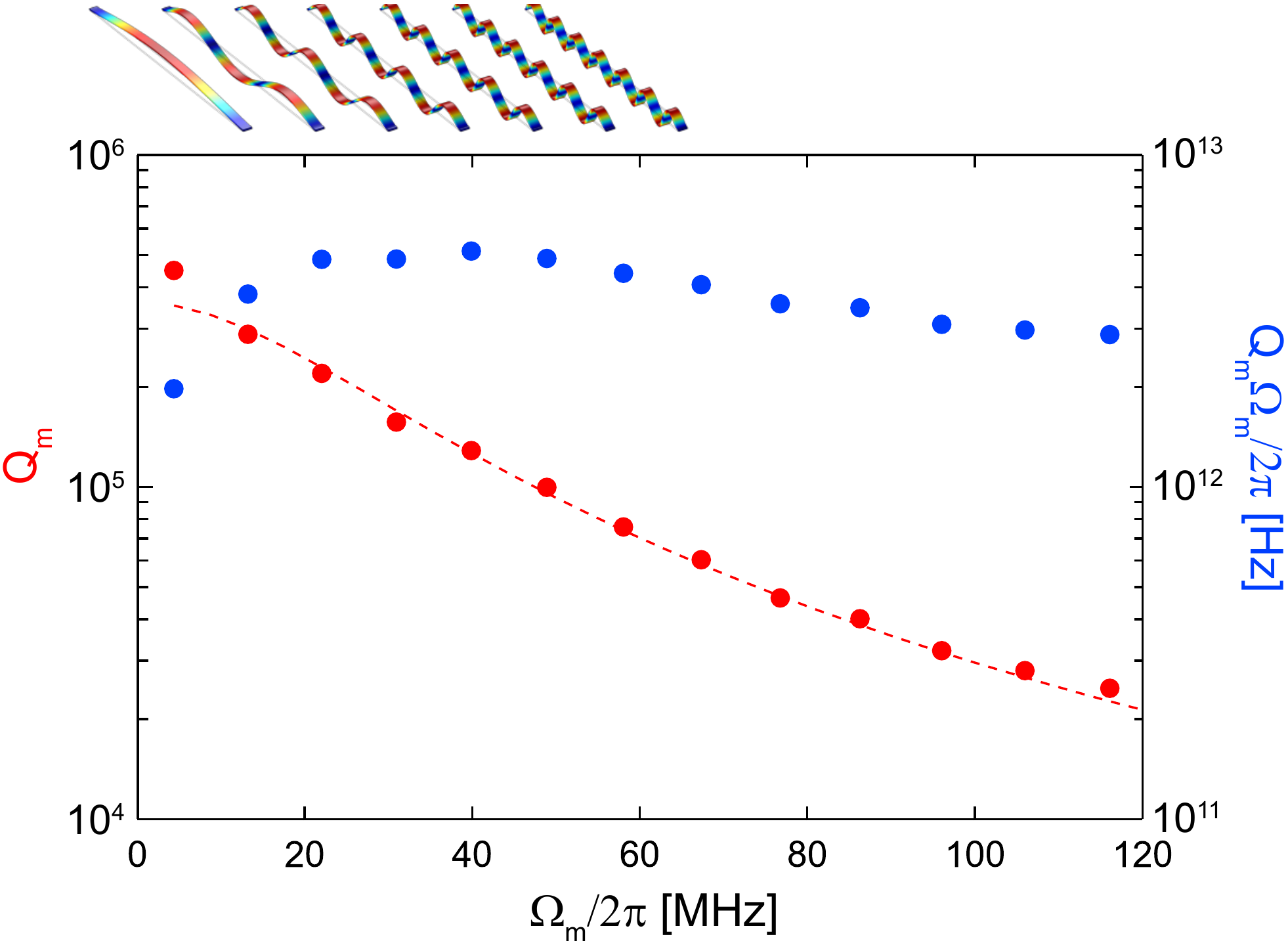}
	\caption{$Q$-factor (red) and $Q\times\t{frequency}$ product (blue) of the first eleven, odd-ordered, out-of-plane flexural modes of a nanobeam with dimensions $\{l,w,t\}=\{60,0.6,0.05\}\;\mu\t{m}$.  Solid red curve is a fit to the $Q$-dilution model in \cite{villanueva2014evidence}, implying a limiting contribution from surface-related intrinsic loss. }\label{fig:design_qf}
	\vspace{0pt}
\end{figure}

In the following sections we carefully detail the design, fabrication, and characterization of the device, and provide a demonstration of low noise displacement measurement.  \secref{sec:design} gives an overview of nanobeam and microdisk resonators and describes a numerical model used to predict their gradient-force optomechanical coupling.  Notably, we find that $G$ can be improved by an order of magnitude by carefully positioning the beam above the disk.  
\secref{sec:fab} describes the fabrication method, particularly the use of planarized (by CMP) sacrificial layers and e-beam lithography, which enable precise engineering of the vertical and horizontal beam-disk separation, respectively. \secref{sec:char} describes characterization of the device using thermomechanical noise measurements and the optical spring effect. In \secref{sec:disp}, we discuss an experiment in which the microdisk is embedded into fiber-based homodyne interferometer, enabling displacement read-out with an imprecision 33 dB below $S^\t{zp}_\omega(\Omega_\t{m})$ for the fundamental beam mode. Finally, in \secref{sec:conclusion}, we remark on the feasibility of Heisenberg-limited position measurements and functionalized applications which take advantage of the heterogeneous integration method.

\section{Device design}\label{sec:design}

\subsection{Nanomechanical beam}\label{sec:nanobeam}

The mechanical resonator we study is a doubly-clamped beam released from a high-stress Si$_3$N$_4$ thin film \cite{verbridge_high_2006}.  Stressed ``nanobeams" are attractive for their string-like flexural modes, which possess exceptionally high $Q/m$ ratios \cite{imboden2014dissipation}.  Beams with of the dimensions studied --- $\{\t{length}\,(l),\t{width}\,(w), \t{thickness}\,(t)\} \sim  \{100, 1, 0.1\}\;\mu\t{m}$ --- possess effective masses $m\sim 10$ pg, fundamental frequencies $\Omega_\t{m}\sim 2\pi\cdot10$ MHz and room temperature quality factors $Q_\t{m}> 10^5$ %and Q-frequency products $Q_\t{m}\cdot\Omega_\t{m}/2\pi>10^{12}$ 
\cite{verbridge_high_2006}.  Significantly, $Q_\t{m}$ is well in excess of the ``universal" value of $10^3-10^4$ observed for bulk amorphous glass resonators at temperatures above $T\gtrsim 1$ K \cite{pohl2002low}. It is also higher than for typical unstressed, single-crystal nanobeams due to surface loss \cite{villanueva2014evidence}. This exceptional behavior is known to derive from a combination of large impedance mismatch from the anchoring body \cite{rieger2014energy} (suppressing extrinsic loss) and stress-related ``dilution" of intrinsic loss \cite{gonzalez1994brownian,unterreithmeier2010damping,villanueva2014evidence}.  From the standpoint of quantum-limited measurement, an important consequence of their high $Q/m$ is that high-stress nanobeams exhibit large zero-point fluctuations.  Expressed as a single-sided spectral density evaluated at the mechanical frequency, the above parameters correspond to a peak zero-point displacement noise density of $S_x^\t{zp}(\Omega_\t{m})= 2\hbar Q_\t{m}/m\Omega_m^2\sim 10\;\t{fm}/\sqrt{\t{Hz}}$.  This value occurs in a radio frequency window, 1-10 MHz, where low noise electronics and laser sources are available; as such, nanobeams were the first solid state mechanical resonators to be read out electrically (using a metal beam) \cite{teufel_nanomechanical_2009} and optically \cite{anetsberger_measuring_2010} with an imprecision lower than $S_x^\t{zp}(\Omega_\t{m})$. 

Measurements of $Q_\t{m}$ for a typical disk-integrated beam with dimensions $\{l,w,t\}=$ $\{60,0.6, 0.06\}$ $\mu\t{m}$ are shown in Fig. \ref{fig:design_qf}.  Despite the complexity of the fabrication procedure (\secref{sec:fab}), flexural modes exhibit $Q_\t{m}\cdot\Omega_\t{m}/2\pi$ as high as $4\cdot 10^{12}$ Hz, on par with the state-of-the-art for high-stress Si$_3$N$_4$ nanobeams of similar dimensions \cite{verbridge2008size,villanueva2014evidence}.  The near-linear eigenfrequency spectrum, $\Omega_\t{m}^{(n)}\approx 2\pi n\cdot4.3\;\t{MHz}$, is consistent with a tensile stress of $\mathcal{T}\approx (\rho l\Omega_m^{(0)}/\pi)^2\approx 0.8$ MPa assuming a density of $\rho=2700 \;\t{kg/m}^3$ \cite{verbridge_high_2006}.  The mechanical-Q spectrum, $Q_\t{m}^{(n)}\approx3.6\cdot10^5/(1+0.023\cdot n^2)$, is consistent with the intrinsic loss model of \cite{unterreithmeier2010damping,villanueva2014evidence}.  The dashed line in \fref{fig:design_qf} is a fit to this model: $Q_\t{m}^{(n)}=Q_\t{int}/(\lambda+n^2\pi^2\lambda^2)$, where $\lambda^2 = Et^2/(12\mathcal{T}l^2)$, $E$ is the elastic modulus of the film, and $Q_\t{int}$ is the intrinsic quality factor of the film when unstressed.  The inferred value of $Q_\t{int}\approx 6700$ (using $E=200$ GPa), is roughly an order of magnitude lower than that for bulk Si$_3$N$_4$.  Interpreted as surface loss, however, the inferred coefficient of $Q_\t{int}/t \approx 1.1\cdot10^{5}\;\mu\t{m}^{-1}$ is within a factor of two of the typical value for LPCVD SiN thin films \cite{villanueva2014evidence}.  Operating in a $^3$He cryostat at 0.5 K, we have recently observed $Q_\t{int}>10^4$ \cite{ghadimi__2014}.%, suggesting that surface loss may be ultimately quenched at these temperatures.

In addition to its favorable mechanical properties when stressed, Si$_3$N$_4$ is an attractive optical material.  It has a relatively large index of refraction, $n\approx 2$, and, owing to its $\sim 3$ eV bandgap, respectably low optical absorption at near infrared wavelengths, characterized by an imaginary index of $n_\t{im}\sim 10^{-5}-10^{-6}$ \cite{zwickl_high_2008}. 
 
\subsection{Optical microdisk}\label{sec:microdisk}

\begin{figure}[t]
	\centering
	\vspace{-0pt}
	\includegraphics[width=1\linewidth]{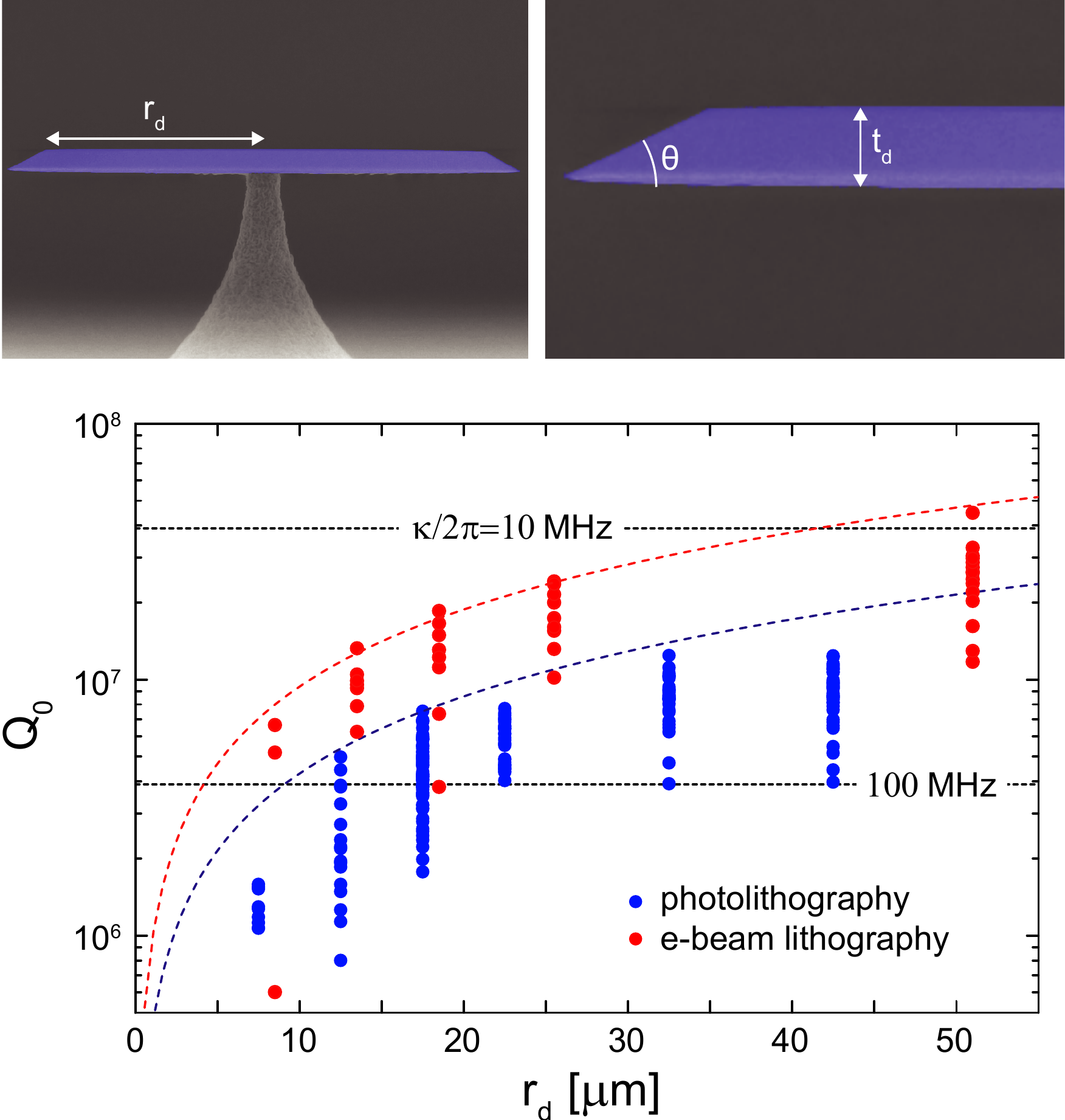}
	\caption{WGM intrinsic quality factor $Q_\t{o}$ as a function of disk radius $r_\t{d}$ for stand-alone SiO$_2$ microdisks of thickness $t_\t{d}\approx 700\,\t{nm}$. TE and TM modes are not distinguished.  Blue (red) points correspond to disks prepared with photolithography (e-beam lithography), which produce wedge angles of $\theta \approx 30(11)^\circ$.   Horizontal lines represent constant cavity linewidth, $\kappa = 2\pi c/(\lambda Q_o)$, with $\lambda = 780$ nm.  Blue (red) dashed line is a guide-to-the-eye for $Q\propto r_\t{d}$, corresponding to a fixed finesse of $\mathcal{F}=0.6\;(1.2)\cdot10^5$. A SEM of a wedged microdisk is shown above; blue (gray) indicates SiO$_2$ (Si).}
	\label{fig:design_kappa}
	\vspace{-10pt}
\end{figure}

The optical resonator we employ is a SiO$_2$ microdisk supporting WGMs along its periphery. SiO$_2$ microdisks possess several advantages for evanescent sensing. The first advantage is that SiO$_2$ exhibits a wide transparency window and a large power handling capacity, enabling large intracavity photon numbers ($n_\t{c}$). The practically achievable $n_\t{c}$ is typically limited by Kerr and Raman nonlinearity. At visible and telecommuncation wavelengths, other effects such as multi-photon absorption do not play a significant role in SiO$_2$, in contrast to Si and other semiconductors.  A second advantage is that standard lithographic techniques, in conjunction with wet-etching, can produce SiO$_2$ microdisks with exceptionally high $Q$ (recently exceeding $10^8$ in the telecommunication band \cite{lee_chemically_2012}). This feature is related to the wedged rim of the disk, which supports WGMs that are spatially isolated from the surface, and thereby from surface scattering/absorption loss. A third advantage is that microdisk WGMs can be evanescently coupled to tapered optical fibers with high ideality \cite{spillane_ideality_2003}. This feature is critical for sensing applications, in which optical loss produces elevated shot-noise imprecision \cite{schliesser_high-sensitivity_2008}.  

\begin{figure*}[ht!]
	
	\centering
	\includegraphics[width=1\linewidth]{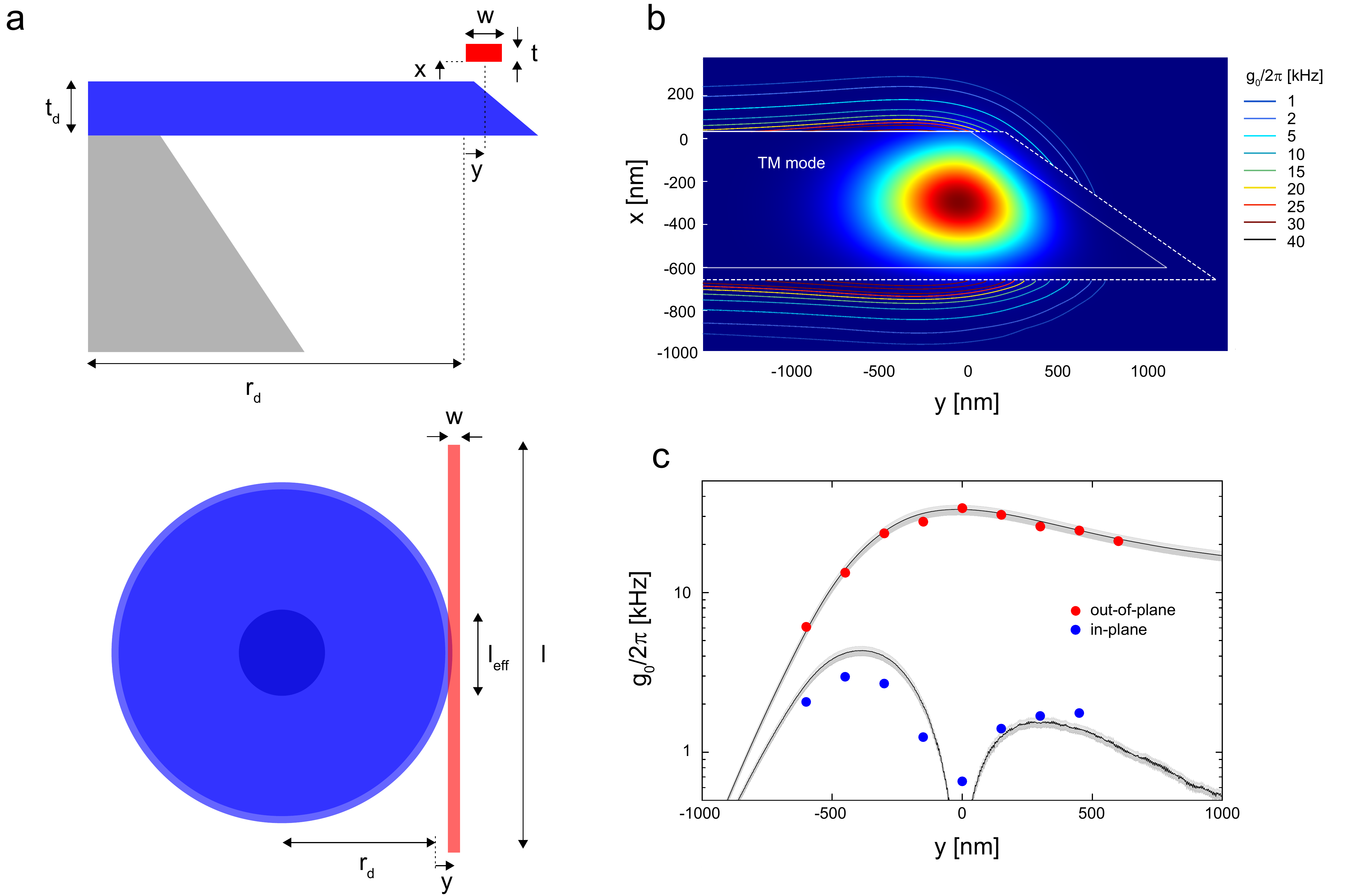}
	\caption{(a) Geometry of the nanobeam-microdisk system: $x$, $y$ represent the vertical (out-of-plane) and lateral (in-plane) position of the beam, respectively, with respect to the inner rim of the disk (thickness $t_\t{d}$, radius $r_\t{d}$). (b) Simulated optomechanical coupling versus beam position for device dimensions $\{t, w, l,x,r_\t{d},t_\t{d}\}= \{0.06, 0.4, 60,0.025,14.2,0.65\}\;\mu\t{m}$.  The intensity profile of a TM-like WGM (computed using FEM) is shown in the background.  Solid and dashed white lines denote the disk surface and the boundary within which the beam touches the disk surface, for the coordinate system defined in (a). Contours indicate lines of constant $g_0$ for the 4.3 MHz fundamental out-of-plane mode. (c) Measured and simulated $g_0$ versus $y$.  Black and blue data are for fundamental out-of-plane and in-plane vibrational modes, respectively (see \secref{sec:char_lateral}). Black lines correspond to numerical solutions to \eqref{eq:G} with a vertical offset of $x=25$ nm.  Gray shading shows the solution space for $x=20-30$ nm.}\label{fig:design_sketch}
\vspace{-5pt}
\end{figure*}

Microdisk resonators were in this case studied at $\lambda\approx700-800$ nm (outside of the telecommunications window), to allow for smaller optical mode volumes. As discussed in \secref{sec:g0model}, reducing the disk radius ($r_\t{d}$) and thickness ($t_\t{d}\sim\lambda/n$) results in smaller mode volumes with fractionally larger evanescent components, thereby increasing the optomechanical coupling strength.  %For disks with thickness $t_\t{d}>300$ nm, numerical simulations reveal that radiation loss remains insignificant for a microdisk radii greater than 10 $\mu$m.
\fref{fig:design_kappa} shows measurements (see \secref{sec:exp} for details) of optical $Q$ versus disk radius ($r_\t{d}$) for microdisk samples of thickness $t_\t{d}=0.7\;\mu\t{m}$ . Two sets of devices are considered. The first set was prepared with photolithography, the second with electron-beam lithography.  The sets differ by their corresponding wedge angle, which is 30 (11) degrees for wet (e-beam) lithography.  For both disk preparation methods, intrinsic $Q>10^6$ was measured for radii as low as $10\;\mu\t{m}$,  corresponding to loss rates of $\kappa\sim2\pi\cdot100$ MHz.  For shallower wedge angles, $Q$ as high as $4\cdot10^7$ ($\kappa\sim2\pi\cdot10$MHz) was obtained --- notably exceeding (for the same $r_\t{d}$) those measured at telecom wavelengths, where scattering losses are significantly lower \cite{kippenberg_demonstration_2006,lee_chemically_2012}.  Numerical simulations \cite{oxborrow2007simulate} reveal that radiation contributes negligibly to the measured loss. Dotted blue (red) lines in \fref{fig:design_kappa} are guide-to-the-eye models for $Q\propto r_\t{d}$, consistent with loss due to surface absorption/scattering \cite{borselli2005beyond}, and corresponding to a fixed finesse of $\mathcal{F}\equiv \Delta\omega_\t{FSR}/\kappa\approx c/(r_\t{d}\kappa)=0.6\;(1.2)\cdot10^5$.  %It should be emphasized that the highest measured quality factors,  $Q\approx4\cdot10^7$, are similar to those measured at telecommunication wavelengths, where scattering losses are significantly lower \cite{kippenberg_demonstration_2006,lee_chemically_2012}.  
As discussed in \secref{sec:char_lateral}, the intrinsic microdisk $Q$ is ultimately reduced by loss introduced by the nanobeam, for beam-disk separations of less than 100 nm.

\subsection{Evanescent optomechanical coupling}\label{sec:g0model}

Optomechanical coupling is achieved by placing the nanobeam near the surface of the microdisk, so that its mid-section occupies the evanescent volume of one of the microdisk WGMs.  When the WGM is excited, the beam experiences a gradient force, $F_\t{opt}$.  The magnitude of this force, and likewise the optomechanical coupling factor $G=\partial\omega_\t{c}/\partial x$, can be derived by computing the work done on the WGM, $-\delta U_\t{cav}$, by a small displacement of the beam, $\delta x$: that is, $F_\t{opt} = -\partial U_\t{cav}/\partial x \approx -G U_\t{cav}/\omega_\t{c}$, where $U_\t{cav}$ is the potential energy stored in the cavity field \cite{povinelli2005evanescent,van_thourhout_optomechanical_2010}.
To first order, it can be shown that \cite{anetsberger_near-field_2009}
\begin{subequations}\label{eq:G}
\begin{align}
G&\approx 
\frac{\omega_\t{c}^{(0)}}{2}\frac{\partial}{\partial x}\left(\frac{\int_{\t{beam}}(\epsilon(\vec{r})-1)|\vec {E}^{(0)}(\vec{r})|^2 d^3r}{\int_\t{disk} \epsilon(\vec{r})|\vec {E}^{(0)}(\vec{r})| ^2 d^3r}\right)
\\ &\approx\frac{\omega_\t{c}^{(0)}}{2}\frac{\partial}{\partial x}\left( \frac{n_\t{SiN}^2-1}{n_{\t{SiO}_{2}}}\frac{|E_\t{max}^{\t{(0,beam)}}|^2}{|E_\t{max}^{\t{(0,disk)}}|^2}\frac{V_\t{beam}}{V_\t{disk}}\right)
\end{align}
\end{subequations}
where $\epsilon(\vec{r})$ is the local relative permittivity, $\vec{E}^{(0)}(\vec{r})$ is the unperturbed cavity field amplitude, and $\int_{\t{beam(disk)}}$ indicates an integral over the volume occupied by the beam (disk).  The simplified expression in \eqref{eq:G}b replaces $\epsilon$ with an index of refraction $n$ and parameterizes each integral in terms of the intensity-weighted volume of the beam (disk), $V_\t{beam(disk)}\equiv\int_\t{beam(disk)}{|E_0|^2 d^3r}/|E_\t{max}^\t{(0,beam(disk))}|^2$, where $E_\t{max}^\t{(0,beam(disk))}$ is the maximum of the unperturbed field within the beam (disk).  

To gain physical insight into \eqref{eq:G}, we consider the configuration shown in \fref{fig:design_sketch}.  Here the beam is placed above the disk, so that it samples the vertical evanescence of a WGM.  For simplicity, the transverse dimensions of the beam are assumed to be much smaller than that of the evanescent field; that is, $w\ll \sqrt{A_\t{WGM}}$ and $t \ll x_\t{ev}$, where $A_\t{WGM}$ is the effective cross-sectional area of the WGM and $x_\t{ev}$ is the exponential decay length of the evanescent field.  In this case $V_\t{beam}$ can be approximated as $twl_\t{eff}$, where $l_\t{eff}<l$ is the intensity-weighted ``sampling length'' of the beam.  Likewise $V_\t{disk}$ can be parameterized as $V_\t{disk}\approx2\pi r_\t{d} A_\t{WGM}$, where $r_\t{d}$ is the physical disk radius.  Assuming the form $|E_\t{max}^\t{(0,beam)}|/|E_\t{max}^\t{(0,disk)}| = \xi e^{-\frac{x+t/2}{x_\t{ev}}}$, neglecting the weak position dependence of $V_\t{beam}$, and assuming the effective mass of a point probe, $m=\rho t w l/2$, the vacuum optomechanical coupling rate can be approximated as
\begin{equation}\label{eq:Gapprox}
g_0\approx \frac{1}{2}
\frac{\omega_\t{c}^{(0)}}{x_\t{ev}}\frac{n_\t{SiN}^2-1}{n_{\t{SiO}_{2}}}\frac{twl_\t{eff}}{2\pi r_\t{d} A_\t{WGM}}\xi e^{-\frac{x+t/2}{x_\t{ev}}}\cdot\sqrt{\frac{\hbar}{\rho twl\Omega_\t{m}}}
\end{equation} 
where $\rho$ is the mass density of the beam. In practice $x_\t{ev},A_\t{WGM},$ and $\xi$ must be determined numerically for a wedged microdisk.  An estimate can be made, however, by assuming the mode shape of a microtoroid WGM with a minor radius of $t_\t{d}/2$ \cite{anetsberger_near-field_2009}. In this case, using $n_{\t{SiO}_2}\approx 1.4$, one has $x_\t{ev}\approx \lambda/(2\pi\sqrt{n_{\t{Si0}_2}^2-1})\approx \lambda/12$, $A_\t{WGM}\approx 0.15 r_\t{d}^{7/12}t_\t{d}^{1/4} \lambda^{7/6}$ and $\xi\approx1.1(\lambda/r_d)^{1/3}$ \cite{anetsberger2010novel}.  Using these formulas, the device geometry $\{t, w, l,x,r_\t{d},t_\t{d}\}= \{0.06, 0.4, 60,0.025,14.2,0.65\}\;\mu\t{m}$, and assuming $\lambda = 780$ nm, $n_{\t{SiN}}=2.0$, $\rho = 2700$ kg/m$^3$, $\Omega_\t{m}=2\pi\cdot 4.3$ MHz, and $l_\t{eff}=10\;\mu$m (see \secref{sec:char_harmonics}), \eqref{eq:Gapprox} predicts that $G\approx 2\pi\cdot 1.0$ GHz/nm, $x_\t{zp} \approx 33 $ fm, and $g_0= 2\pi\cdot 33$ kHz.  As shown in \fref{fig:design_sketch}d, this estimate agrees well with numerically and experimentally determined values.  Notably, \eqref{eq:Gapprox} implies that to achieve large $g_0$, it is necessary to reduce the vertical gap to $x<x_\t{ev}\approx 100$ nm, and to maximize $l_\t{eff}$ by laterally positioning the beam \emph{above} the disk.

A numerical model for $g_0(x,y)$ is shown in \fref{fig:design_sketch}b.  Intrinsic WGM mode shapes, $\vec{E}^{0)}(\vec{r})$, were computed using an axially-symmetric finite element model (COMSOL FEM axial symmetric package \cite{oxborrow2007simulate}). The energy stored in the WGM,  $U_\t{cav}^{(0)}\approx\tfrac{1}{2}\int_{\t{disk}}\epsilon(\vec{r})|\vec {E}^{(0)}(\vec{r})|^2 d^3r$, and the energy shift due to the beam, $\Delta U_\t{cav}(x,y)\approx\tfrac{1}{4}\int_{\t{beam}}(\epsilon(\vec{r})-1)|\vec {E}^{(0)}(\vec{r})|^2 d^3r$, were computed by numerical integration in Matlab.  Differentiating the 2D energy landscape gives $G(x,y)=\omega_\t{c}\tfrac{\partial}{\partial x}(\Delta U_\t{cav}(x,y)/U_\t{cav}^{(0)})$ for out-plane-motion. %The transverse mode profile of a TM-like whispering gallery mode is shown in \fref{fig:design_g0model}a, for a disk with dimensions $\{r_\t{d},\theta_\t{d}\}=\{14.2\;\mu\t{m},31\;\t{deg.}\}$.  
\fref{fig:design_sketch}b shows $g_0(x,y)=G(x,y)\cdot x_\t{zp}$ for a beam and disk with the dimensions given above, for a TM-like WGM mode.  %The effective mass is again approximated by $m \approx \rho t l w/2$ (see \secref{sec:char_harmonics}).  
Contours indicate that the optimal position of the beam is above and inside the inner rim of the disk, and that the magnitude of $g_0$ scales exponentially with vertical displacement from the disk surface, with a decay length of $\sim 100$ nm.  A horizontal cut through the contours for $x = 25$ nm is shown in \fref{fig:design_sketch}c.  Upper and lower curves show models for fundamental in-plane (IP) and out-of-plane (OP) flexural modes. Significantly, maximizing $g_0^\t{(OP)}$ also minimizes $g_0^\t{(IP)}$; this opens a wide spectral window, $\Delta\Omega\sim\Omega_\t{m}$, for measurement of the out-of-plane mode.  Experimental measurements (see \secref{sec:TNS}) of $g_0(25\;\t{nm},y)$ are also shown in \fref{fig:design_sketch}c .  The model agrees well with experiment assuming a vertical offset of $25\pm5$ nm.

%\vfill \pagebreak

\section{Device fabrication}\label{sec:fab}

\begin{figure}[t!]
	\vspace{-0pt}
	\centering
	\includegraphics[scale=0.41]{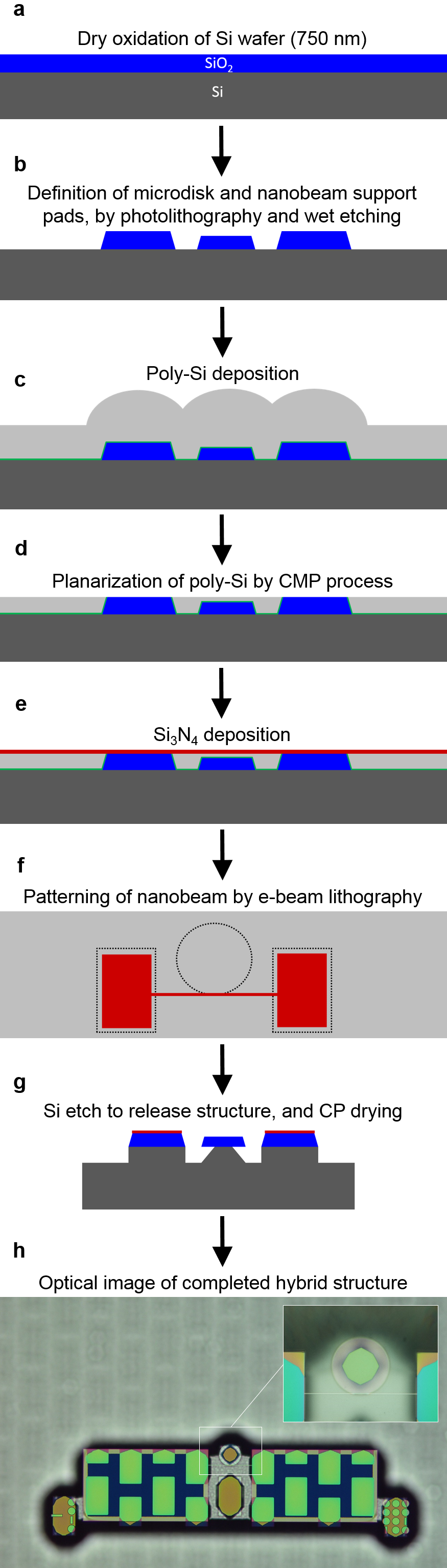}
	\caption{\label{fig:fab_flow}Fabrication process flow: blue, red, green, and (light) gray indicate SiO$_2$, Si$_3$N$_4$, Al$_2$O$_3$, and (poly-)Si, respectively.}
	\vspace{-10pt}
\end{figure}

The fabrication process is outlined in \fref{fig:fab_flow}.  Four key elements of the process, detailed in the following subsections, are: (A) fabrication of the SiO$_2$ microdisk, (B) formation of a planarized sacrificial layer using chemical mechanical polishing (CMP), (C) fabrication of the Si$_3$N$_4$ nanobeam, and (D) release of the sacrificial layer. Of particular importance is the sacrificial layer, which allows the mechanical (Si$_3$N$_4$) and optical (SiO$_2$) elements to be designed independently while maintaining the high optical quality and achieving a vertical beam-disk separation of less than 100 nm. Also important is the use of e-beam lithography to pattern the Si$_3$N$_4$, as this enables fine tuning of the lateral beam position. %We The process has also been engineered to  leaves the door open to further vertical integration. 

\subsection{Microdisk fabrication}\label{sec:fab_microdisk}

\begin{figure}[b!]
	\vspace{-5pt}
	\centering
	\includegraphics[width=.99\linewidth]{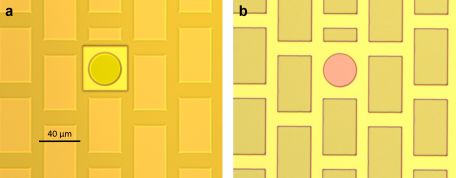}
	\caption{\label{fig:fab_disk_offset} Defining the vertical gap between the disk and the nanobeam: (a) Top view of patterned SiO$_2$ prior to selective etch of the microdisk. Photoresist protects the sacrificial structures, while a window is exposed around the microdisk. (b) Top view after selective etch of the microdisk and removal of the photoresist. The altered color of the microdisk indicates thinning.}
	\vspace{-0pt}
\end{figure}

The process begins with an undoped, float-zone (FZ) Si wafer, on which a 750 nm film of SiO$_2$ is grown by dry oxidation.  Three structures are patterned into the SiO$_2$ film: the microdisk, rectangular pads that later serve as a support for the nanobeam and a reference plane for CMP polishing, and markers that are later used for e-beam alignment. As illustrated in \fref{fig:fab_disk_offset}, the SiO$_2$ pattern is processed in two stages.  In the first stage all structures are defined.  In the second stage the microdisk is etched preferentially, recessing it from the pads and defining the vertical gap between disk and the beam.

Details of the SiO$_2$ patterning process are as follows:  The first mask, containing all structures, is exposed in $1.1\; \mu\t{m}$ of Microchemicals AZ 1512 photoresist using a Karl S\"{u}ss MA 150 mask aligner and a
broadband Hg lamp. A subsequent reflow step is used to smooth the pattern boundaries and minimize standing wave patterns. %The sample is then exposed to a low power O$_2$ plasma for descumming. 
Afterwards, the pattern is transfered to SiO$_2$ by etching in a room-temperature bath of BHF. The photoresist is then stripped and a second mask is applied. The second mask covers all structures on the wafer except for the microdisk, leaving it exposed for etching (\fref{fig:fab_disk_offset}a). %After a second descum, 
The microdisk is preferentially etched in BHF until it is 10-100 nm thinner than the surrounding pads (later defining the beam-disk gap). The result, after the photoresist is stripped, is shown in \fref{fig:fab_disk_offset}b. Note that the microdisk reflects a different color than the surrounding pads due to its reduced thickness.  Also seen in \fref{fig:fab_disk_offset} is a matrix of sacrificial pads surrounding the disk.  This matrix extends across the entire wafer and is only broken where microdisks or alignment marks (not shown) are placed. As discussed in \secref{sec:fab_sacrificial}, a uniform matrix of pads is necessary to achieve a flat surface when performing CMP the sacrificial layer.

The final result of microdisk fabrication is illustrated in \fref{fig:fab_flow}b. Blue indicates (in profile) the patterned SiO$_2$ film, with the microdisk in the center and nanobeam support pads on either side. Not shown are sacrificial pillars and alignment marks. In the next processing step, all structures are buried in a sacrificial material, onto which a Si$_3$N$_4$ film will be grown. 

\subsection{Planarized sacrificial layer}\label{sec:fab_sacrificial}

After patterning, the SiO$_2$ film is covered with a layer of sacrificial material.  The sacrificial layer is used as a substrate for deposition and patterning of the Si$_3$N$_4$ film, meanwhile protecting the underlying microdisk.  A crucial consideration is the thickness and flatness of the sacrifical layer, which is initially uneven because of its conformity to the underlying SiO$_2$ pattern. To thin and planarize the sacrificial layer, a delicate chemical-mechanical polishing (CMP) procedure is followed.

Poly-Si is chosen as the sacrificial material because it can be isotropically etched with high selectivity to SiO$_2$ and Si$_3$N$_4$, is well-suited to CMP, can withstand the high temperatures required for LPCVD Si$_3$N$_4$ ($>800\,^{\circ}\t{C}$), and can be used to undercut the nanobeam and the microdisk in a single step.  A 1.5 $\mu$m thick layer is deposited by LPCVD at $600\,^{\circ}\t{C}$ using silane and disilane as reactants. In addition, immediately before poly-Si deposition, a 5 nm aluminum oxide (Al$_2$O$_3$) film is deposited atop the SiO$_2$ using atomic layer deposition.  This film later serves as an etch-stop to protect the microdisk when releasing the Si$_3$N$_4$ nanobeam. (Al$_2$O$_3$ etches over $100\times$ slower than Si$_3$N$_4$ in flourine-based RIE used, and thus a few nanometers is sufficient to protect the microdisk.)

A profile of the per-polished sacrificial layer is sketched in \fref{fig:fab_flow}c.  The Al$_2$O$_3$ etch-stop film is indicated by green. %ALD achieves atomic layer control of film growth by separating the reactants into 2 precursors that are introduced to the chamber sequentially and cyclically, allowing growth of one molecular layer at a time. This process is used to produce very thin continuous films with high conformality - both of which are critically important here.
Immediately above the etch-stop is the layer of poly-Si (gray).  Because of the underlying SiO$_2$ structures, the surface of the poly-Si is uneven.   This surface is planarized by CMP before Si$_3$N$_4$ is deposited.   

CMP involves pressing the wafer against a rotating polishing pad in the presence of an abrasive and corrosive chemical slurry. Abrasion is provided by SiO$_2$ particles 30-50 nm in diameter. 
The slurry PH is adjusted to achieve the desired polishing rate. In practice the polishing rate is also a function of applied force, rotation speed, and wafer topography. Areas of the wafer where features are sparse experience a higher pressure and thus a higher polishing rate than areas where features are dense. In order to reduce the poly-Si thickness to less than 100 nm over the entire 100 mm wafer, a uniform polishing rate is critical. This is the reason for patterning a matrix of sacrificial pads (\fref{fig:fab_disk_offset}). 

The objective of the CMP process is to remove poly-Si until the pads are exposed, while maintaining a thin layer above the recessed microdisk (\fref{fig:fab_flow}d). This procedure is complicated by the fact that the polishing rate varies across the wafer and, more importantly, that the polishing rate above the microdisk is faster than the rate above the adjacent nanobeam support pads.  The latter results in a poly-Si layer which is thinner above the microdisk than at the nanobeam supports. To reduce this ``dishing'' effect, the support pads are brought as close the microdisk as possible (limited to $7\;\mu\t{m}$ by photolithography and BHF biasing). To further reduce dishing, a two step polishing technique is used. First a slurry designed to etch poly-Si is used to remove the bulk of the material, leaving approximately 100 nm above the pads. The remaining material is removed with a slurry designed to etch SiO$_2$ faster than poly-Si.  When the surface of the SiO$_2$ pads is reached, the dishing effect begins to reverse, resulting in an overall flat surface.

The gap between the microdisk and nanobeam is not determined by the thickness of the sacrificial layer, but rather by the predefined difference in thickness between the microdisk and the pads (\fref{fig:fab_flow}b). During the final steps of CMP, however, the support pads are etched.  The final gap is therefore smaller than originally defined by thinning of the microdisk. In order to precisely tune the gap, the thickness of the clamping pads is iteratively measured by reflectometry until a desired value is reached. The sample is then ready for Si$_3$N$_4$ deposition.

\subsection{Nanobeam fabrication}\label{sec:fab_mechanical}

To form the nanobeam, a 50-100 nm film of high-stress Si$_3$N$_4$ is deposited onto the planarized poly-Si layer. LPCVD is performed at $800\,^{\circ}\mathrm{C}$ using dichlorosilane and ammonia, producing a nearly stoichiometric Si$_3$N$_4$. High stoichiometry is important for reducing absorption caused by hydrogen and oxygen impurities \cite{zwickl_high_2008}. Moreover, the stress (800 MPa) resulting from high temperature deposition is important for achieving high mechanical quality factors \cite{verbridge_high_2006}.

%As shown in sections \ref{sec:simulation} and \ref{sec:char_lateral}, 
To maximize optomechanical coupling, it is necessary to fine tune the lateral beam-disk separation with $100$ nm precision (\fref{fig:design_sketch}c). %In fact, moving the nanobeam by 150 nm is enough to change the single-photon optomechanical coupling rate by approximately two, as can be seen in figure \fref{fig:char_lateral}. Thus it is important to have precise alignment to the microdisk. 
This is accomplished using e-beam lithography to define the beams, in conjunction with the alignment markers defined during SiO$_2$ patterning. Importantly, after Si$_3$N$_4$ deposition, the markers are buried under Si$_3$N$_4$ and poly-Si, and cannot be seen by the electron-beam.  A series of etch steps are used to locally uncover the markers; in addition, to improve contrast, the exposed markers are used as a hard mask to etch $2\;\mu\t{m}$ into the underlying Si, using a highly selective flourine-based etch. The resulting high-contrast markers permit alignment of the Si$_3$N$_4$ mask with sub-100 nm precision.

The nanobeams, support pads, and sample labels are patterned in a 180 nm-thick hydrogen silsesquioxane (HSQ) negative photoresist. %After development in tetramethylammonium hydroxide, HSQ is chemically similar to SiO$_2$. 
To reduce writing time, the pattern is separated into two parts, one containing the nanobeams and one containing the pads and labels. The former is written with a high resolution of 5 nm, while the latter is written with a 50 nm resolution. Proximity effect correction is used to ensure a high fidelity pattern. %Proximity effect correction software calculates the dose from this backscattering at each grid point and adjusts the writing dose to ensure the correct \emph{effective dose} is achieved.
The e-beam pattern is transferred to Si$_3$N$_4$ using an SF$_6$ RIE etch. The resulting structure is shown in \fref{fig:fab_sample_pattern}a. 

\begin{figure}[b]
	\vspace{-5pt}
	\centering
	\includegraphics[width=.99\linewidth]{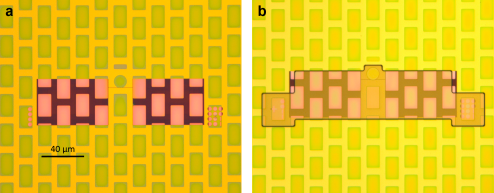}
	\caption{\label{fig:fab_sample_pattern} Defining the nanobeam and the "mesa". (a) Top view of sample after etching of Si$_3$N$_4$ (pink and purple). Surrounding SiO$_2$ structures, including microdisk, appear green. (b) Image of the ``mesa" photomask.}
	\vspace{-0pt}
\end{figure}

\subsection{Structural release}\label{sec:fab_release}

\subsubsection{Mesa and sample chip}
Before the nanobeam and microdisk are released, they are elevated from the surrounding wafer on a rectangular ``mesa''. This later enables alignment of a straight tapered optical fiber to the microdisk \cite{riviere_evanescent_2013}.  Figure \ref{fig:fab_sample_pattern}b shows the mesa defined in a $5\;\mu\t{m}$ mask of Microchemicals AZ 9260 photoresist. Flourine-based RIE is used to remove the surrounding poly-Si.  The underlying sacrificial SiO$_2$ pads are removed by a subsequent BHF etch, exposing the Si substrate. To create the elevated mesa, exposed Si is recessed an additional $50\;\mu\t{m}$ by DRIE. 

After releasing the mesa, the sample chips are defined.  To define the sample chips, the wafer is coated with a protective photoresist layer and partially diced ($300\;\mu\t{m}$ deep) with a high precision Si dicing saw. Partial dicing is important as it leaves the wafer intact, enabling further processing using wafer-scale equipment. After partial dicing the photoresist is stripped, so that final release steps can be carried out. 

\subsubsection{Nanobeam and microdisk}

To release the nanobeam and undercut the microdisk, the partially diced wafer is immersed in $40\%$ KOH at $45\,^{\circ}\mathrm{C}$, selectively removing poly-Si but also etching Si. The etch time is fine-tuned with two opposing criteria in mind: (1) to ensure that the microdisk is undercut sufficiently far from its rim to avoid optical losses and (2) to ensure that Si underneath the nanobeam clamping point is not etched away.  After KOH etching, the wafer is rinsed in water and any remaining potassium is neutralized in a bath of hydrochloric acid.  Organic cleaning is then performed using an exothermic mixture of three parts sulfuric acid to one part 30\% hydrogen peroxide (a ``piranha etch''). After rinsing again, the wafer is transfered directly to the ethanol bath of a critical point drying (CPD) machine. %CPD is a technique used to dry suspended parts that would otherwise stick together under the tension of evaporative drying. This is accomplished through avoiding the liquid to gas phase transition and instead passing through the supercritical regime. CPD does this by replacing EtOH with carbon dioxide ($CO_2$), and then controlling the pressure and temperature of the $CO_2$, such that the transition from liquid to gas is circumvented via the supercritical regime. 
%(We note that successful suspension of the nanobeams could not be achieved without CPD.) 
After CPD, the wafer is broken into sample chips along the partially diced lines, concluding the fabrication process.

\section{Device characterization}\label{sec:char}

\begin{figure*}[ht!]
	\centering
	\includegraphics[width=0.95\linewidth]{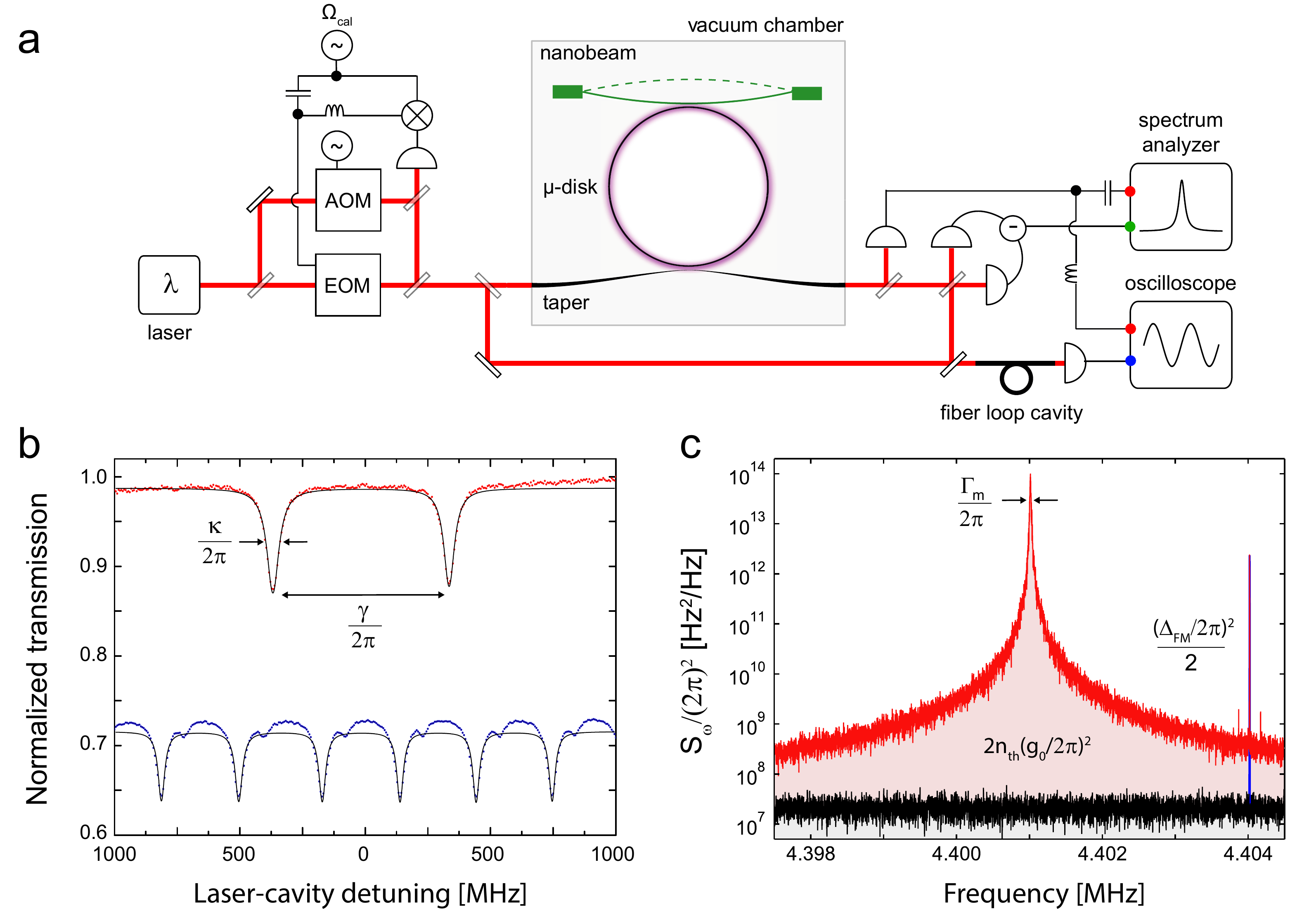}
	\caption{(a) Overview of the experimental apparatus, described in  \secref{sec:exp}. (b) Representative optical Q measurement.  WGM loss rates ($\kappa$) and mode splitting ($\gamma$)  are inferred from the cavity transmission profile (red), generated by sweeping the diode laser frequency while monitoring the transmitted power.  The sweep is calibrated by simultaneously monitoring transmission through a fiber loop cavity (blue). (c) Representative thermomechanical noise measurement.  $\Omega_\t{m},\Gamma_\t{th}$, and $g_0$ are inferred from the center frequency, linewidth, and area beneath the thermal noise peak (pink), respectively.  The latter is calibrated by normalizing to the area beneath a FM tone (blue).}\label{fig:char_experimentalsetup} 
\end{figure*}

\subsection{Experimental setup}\label{sec:exp}

Samples are characterized using the experimental setup shown in \fref{fig:char_experimentalsetup}. Light from a $765-785$ nm tunable diode laser (New Focus Velocity 6312) is coupled into the microdisk using a tapered optical fiber (780 HP) \cite{spillane_ideality_2003}.  The forward-scattered (``transmitted") field is monitored using one of two techniques: direct detection with an avalanche photodiode (Thorlabs APD110) and balanced homodyne detection with a pair of fast Si photodiodes (FEMTO HCA-S-100).  DC- and AC-filtered photosignals are split between an oscilloscope (Tektronix DPO4034) and a spectrum analyzer (Tektronix RSA5106A).  To calibrate laser-cavity detuning, a fraction of the input field is simultaneously passed through a $20$-cm-long (FSR $\sim350$ MHz) fiber loop cavity.  Optical decay rates are inferred from measurements of transmitted power versus laser detuning (\fref{fig:char_experimentalsetup}b).  Mechanical properties, including the optomechanical coupling rates, are inferred from measurements of thermomechanical cavity frequency noise \cite{gorodetsky_determination_2010} (\fref{fig:char_experimentalsetup}c).  To calibrate this noise, the input field is frequency modulated using an electro-optic modulator (EOSpace).  Residual amplitude modulation --- an important source of calibration error --- is actively suppressed by stabilizing the phase of an out-of-loop heterodyne beat \cite{zhang2014reduction}.  To eliminate gas damping of the nanobeam, the sample chip and the fiber coupling setup (based on an Attocube stack) are embedded in a vacuum chamber operating at $<10^{-5}$ mbar.

%\subsection{Cavity transmission spectroscopy}\label{sec:char:optics}
%
%Optical decay rates ($\kappa$) in \fref{fig:design_kappa} were inferred by cavity transmission spectroscopy: that is, by monitoring the power transmitted through the taper versus laser-cavity detuning.  Measurements were made with the taper positioned so that the external coupling rate, $\kappa_\t{ex}$, was small ($<10\%$) compared to the internal loss rate $\kappa_0 = \kappa-\kappa_\t{ex}$ \cite{spillane_ideality_2003}.  The power of the input field was attenuated to a level weak enough to avoid thermal nonlinearities (typically $\sim 100$ nW).  The laser frequency was then swept over the WGM resonance at a rate $\sim 1$ THz/sec.  A representative trace is shown in \fref{fig:char_experimentalsetup}b. The horizontal axis was calibrated by simultaneously monitoring the transmission of the laser field through the fiber loop cavity.  $\kappa$ and modal splitting $\gamma$ \cite{kippenberg_modal_2002} are estimated by fitting the trace to a double-Lorenztian. 

\subsection{Thermal noise measurement}\label{sec:TNS}

Mechanical mode frequencies $\Omega_\t{m}$, damping rates  $\Gamma_\t{m}$, and optomechanical coupling rates $g_0$, were determined by analyzing the cavity resonance frequency noise produced by thermal motion of the nanobeam. An in-depth description of this method is given in \cite{gorodetsky_determination_2010}.  Important details are recounted below for clarity.

Thermal motion of the nanobeam $x(t)$ is written onto the cavity resonance frequency $\omega_c(t)$ via their optomechanical coupling, $G=d\omega_c/dx$. To measure $\omega_c(t)$, we monitor the power of the transmitted field while operating at a fixed detuning of $|\Delta|\approx \kappa/2$.  Referred to the output voltage ($V$) of the photodetector transimpedance amplifier, the uncalibrated noise spectrum 
can be expressed as $S_V(\Omega)=|G_{V\omega}(\Omega)|^2 S_\omega(\Omega)$, where $G_{V\omega}(\Omega)$ is the measurement transfer function and $S_\omega(\Omega)$ is the apparent cavity frequency noise. $G_{V\omega}(\Omega)$ is calibrated by applying a phase modulation tone of known depth ($\beta_\t{cal}$) and frequency ($\Omega_\t{cal}$) to the input, resulting in a narrow spectral peak with area $|G_{V\omega}(\Omega_\t{cal})|^2\beta_\t{cal}^2\Omega_\t{cal}^2/2$ \cite{gorodetsky_determination_2010}.

A representative measurement is shown in  \fref{fig:char_experimentalsetup}c. Red, blue, and black components correspond to thermal noise, $S_\omega^\t{th}(\Omega)$, the calibration tone, $S_\omega^\t{cal}(\Omega)$, and measurement imprecision, $S_\omega^\t{imp}(\Omega)$, respectively.  The full signal can be modeled as
\begin{subequations}\label{eq:Somega}
\begin{align}
S_{\omega}(\Omega)&=S^\t{th}_{\omega}(\Omega) + S^\t{cal}_{\omega}(\Omega) + S_{\omega}^\t{imp}(\Omega) \\ &\approx 2g_0^2n_\t{th}\cdot \mathcal{L}(\Omega-\Omega_\t{m})\\&\;\;\;\;+\frac{\beta_\t{cal}^2\Omega_\t{cal}^2}{2}\cdot \mathcal{G}(\Omega-\Omega_\t{cal})+S_{\omega}^\t{imp}(\Omega),
\end{align}
\end{subequations}
where  $\mathcal{L}(\Omega)=4\Gamma_\t{m}/(\Gamma_\t{m}^2+4\Omega^2)$ is a normalized Lorentzian (characterizing the mechanical susceptibility) and $\mathcal{G}(\Omega)=e^{-\Omega^2/(2B^2)}/\sqrt{2\pi B^2}$ is a normalized Gaussian (characterizing the window function of the spectrum analyzer, which is assumed to have a resolution bandwidth $B\ll\Gamma_\t{m}$).  To calibrate the vertical axis in \fref{fig:char_experimentalsetup}c, it is assumed that $|G_{V\omega}(\Omega_\t{m})|\approx |G_{V\omega}(\Omega_\t{cal})|$.  
Fitting the calibrated spectrum %, $S_\omega(\Omega)\approx S_V(\Omega)/|G_{V\omega}(\Omega_\t{cal})|^2$
to \eqref{eq:Somega} gives $\Omega_\t{m}$, $\Gamma_\t{m}$, and $g_0$.  The last inference requires knowledge of $n_\t{th}$.  By using input powers low enough to neglect photothermal/radiation pressure damping ($\sim 10$ nW), we assume that $n_\t{th}\approx k_\t{B}\cdot 295\;\t{K}/(\hbar\Omega_\t{m})$. 

\subsection{Optical spring effect}

\begin{figure}[t]
	\centering
	\includegraphics[width=1\linewidth]{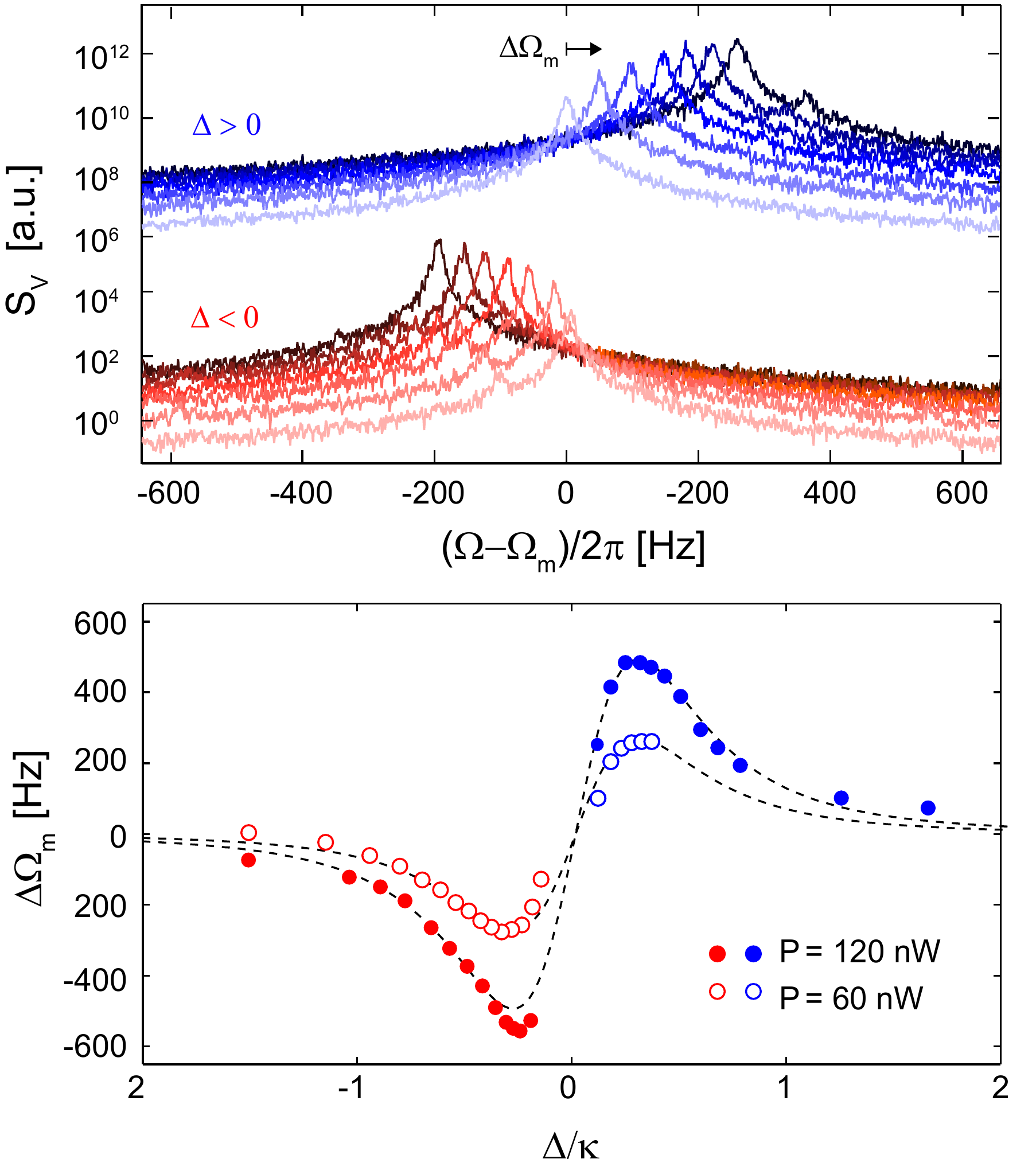}
	\caption{Optical spring measurement.  (a) Thermal noise spectrum of the fundamental beam mode as a function of laser detuning.  Blue and red spectra indicated blue ($\Delta>0$) and red ($\Delta<0$) detuning, respectively.  Lighter shades indicate smaller detuning.  Blue spectra are vertically offset.  (b) Plot of optical spring shift, $\Delta\Omega_\t{m}$, versus normalized detuning, $\Delta/\kappa$.  Dashed black lines are a fit to \eqref{eq:opticalspring} using $g_0$ as a free parameter.}
	\label{fig:spring}
	\vspace{-10pt}
\end{figure}

As a cross-check of the thermal noise measurement, $g_0$ was independently estimated from the optical spring effect \cite{aspelmeyer_cavity_2015}.  In the experimentally relevant bad cavity limit ($\Omega_\t{m}\ll \kappa$), the mechanical frequency shift produced by a radiation pressure optical spring is
\begin{equation}
\Delta \Omega_\t{m}(\Delta)\approx \frac{8g_0^2}{\kappa}\cdot n_\t{c}(\Delta)\cdot \frac{\Delta/\kappa}{1+4(\Delta/\kappa)^2}%\le  \frac{g_0^2}{\kappa}\cdot n_\t{c}(0)\cdot \frac{3\sqrt{3}}{4}
\label{eq:opticalspring}
\end{equation}
where $\Delta$ is the laser-cavity detuning, $n_\t{c}(\Delta)= (4P/(\hbar\omega_\t{0}\kappa))(\kappa_\t{ex}/\kappa)/(1+4(\Delta/\kappa)^2)$ is the intracavity photon number, and $P$ is the power injected into the cavity.  (We note that radiation pressure damping also occurs for a detuned input field; however, in the devices studied, for which $\Omega_\t{m}/\kappa\sim 0.01$, this effect was found to be overwhelmed by photothermal damping \cite{metzger2004cavity}.) 

A measurement of the optical spring effect is shown in \fref{fig:spring}, corresponding to the sample also characterized in \fref{fig:char_experimentalsetup}c.  The injected powers used --- $P = 60,120$ nW --- were chosen to avoid instabilities due to photothermal/radiation pressure damping.  The cavity was critically coupled ($\kappa_\t{ex}\approx\kappa/2\approx 2\pi\cdot 550$ MHz) and laser detuning was estimated from the mean transmitted power.  Overlaid models correspond to \eqref{eq:opticalspring} with the value $g_0= 2\pi\cdot 60$ kHz, inferred from a least-squared fit to the low power measurement.  This value is within 10$\%$ of that inferred from thermal noise in \fref{fig:char_experimentalsetup}c.

\begin{figure*}[t!]
	\centering
	\includegraphics[width=1\linewidth]{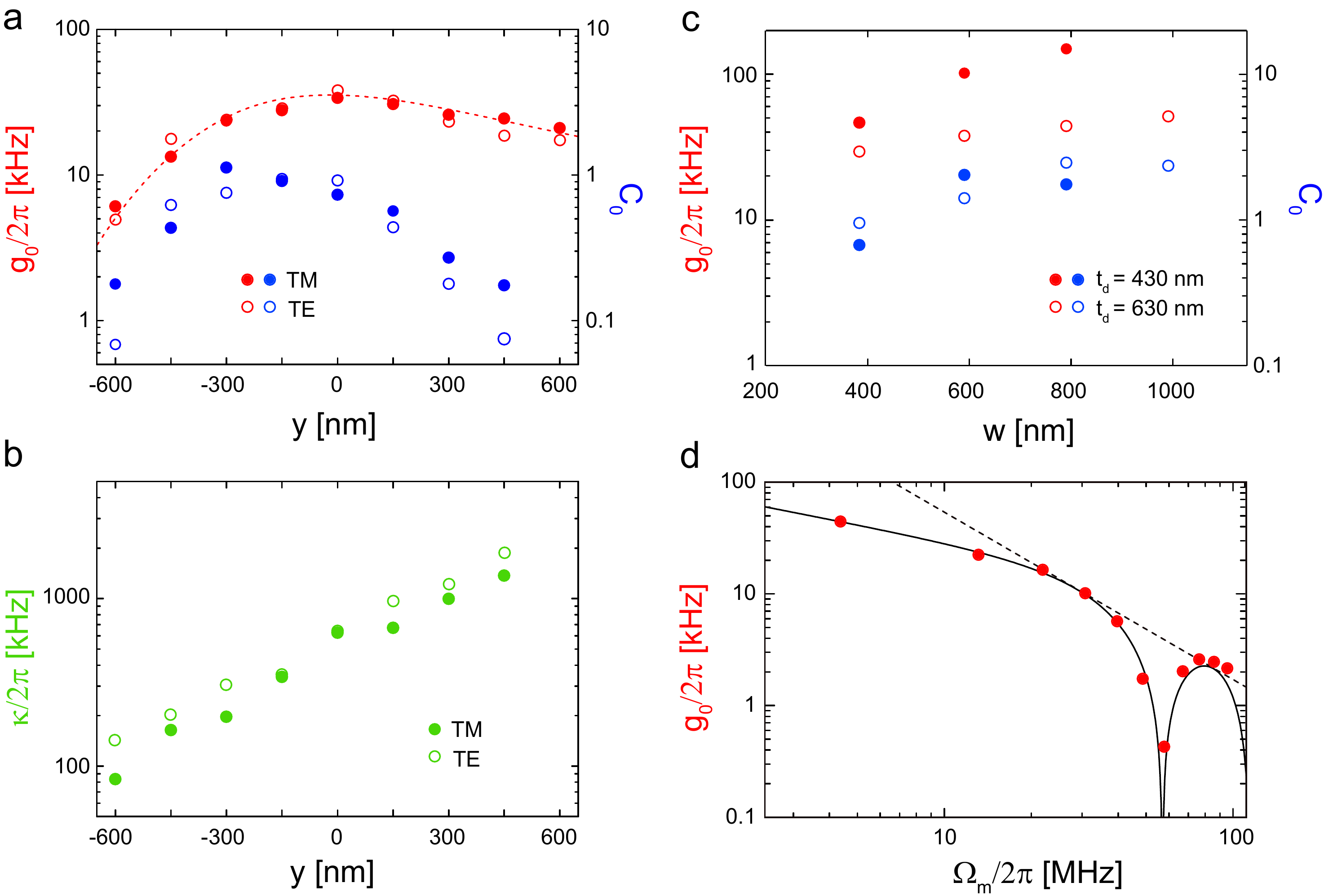}
	\caption{(a) Measured vacuum optomechanical coupling rate ($g_0$) and cooperativity ($\mathcal{C}_0$, assuming $\Gamma_\t{m}=2\pi\cdot15$ Hz) versus lateral beam position ($y$) for TM (solid circles) and TE (open circles) cavity modes. (b) Corresponding intrinsic cavity decay rate ($\kappa_0$). (c)  Measured $g_0$ versus beam width ($w$) for two disk thicknesses ($t_\t{d}$). (d) Measured $g_0$ versus mode frequency, $\Omega_\t{m}^{(n)}\approx n\Omega_\t{m}^{(0)}$.  Red dots correspond to odd harmonics ($n = 1,3,5...$).  Solid and dashed lines are model curves (Eq. \ref{eq:g0vsn}b) for a sampling length of $l_\t{eff}=9.6$ and $l_\t{eff} = 0$, respectively}%, using $g_0^{(n)}/g_0^{(0)}\approx \lvert\t{sinc}\left(\frac{n \pi}{2}\frac{l_\t{eff}}{l}\right)\rvert/\sqrt{n}$}
	\label{fig:char_lateral}
	\vspace{0pt}
\end{figure*}

\subsection{$g_0$ and $\mathcal{C}_0$ versus lateral beam position}\label{sec:char_lateral}

As discussed in \secref{sec:g0model}, $g_0$ depends sensitively on the lateral positioning of the nanobeam, and assumes a maximum (minimum) value for out-of-plane (in-plane) flexural modes when centered above the WGM.  This behavior was studied by sweeping the lateral position of the beam using an appropriate e-beam mask (\secref{sec:fab_mechanical}). Measurements of $g_0$ versus lateral beam position are shown in \fref{fig:char_lateral}a.  (In-plane modes exhibit typically $10\times$ lower $g_0$, and were not considered.) for beam and disk dimensions of $\{l,w,t\}=\{60,0.4,0.06\}\;\mu$m and $\{r,t_\t{d},\theta\}=\{15\,\mu\t{m},0.60\,\mu\t{m},30\,\t{deg.}\}$, respectively, and for a vertical gap of 25 nm.  In agreement with numerical modeling (dashed line), $g_0$ assumes a maximum of $2\pi\cdot 40$ kHz as the outer edge of the beam eclipses the rim of the disk.  %Notably, the observed $g_0>2\pi\cdot 10$ kHz is 20 dB larger than our previous chip-scale device \cite{gavartin_hybrid_2012}, owing to the small vertical gap and optimal lateral placement of the beam.

Also shown in \fref{fig:char_lateral}b are measurements of $\kappa$ versus lateral beam position ($y$).  When the beam is displaced far from the disk, $\kappa$ converges to the intrinsic value of $\sim 2\pi\cdot 100$ MHz observed in \fref{fig:design_kappa}, suggesting that CMP did not significantly affect microdisk surface quality. As the beam is brought within 100 nm of the disk, $\kappa$ is observed to increase sharply.  The observed exponential dependence $\kappa$ on $y$ is independent of mode polarization and similar to the scaling observed in \cite{anetsberger_measuring_2010} with a beam coupled to a microtoroid. % We remark that the similar behavior for TE and TM optical modes is inconsistent with loss due to waveguide coupling to the beam.  
The absolute magnitude of the loss is also inconsistent with bulk Si$_3$N$_4$ optical absorption --- specifically, accounting for the relatively small fraction of energy stored in the beam, the observed loss would require an imaginary index of $\sim10^{-4}$, which is 1-2 orders of magnitude larger than conventionally observed for Si$_3$N$_4$ at NIR wavelengths \cite{zwickl_high_2008,wilson2009cavity}. We thus conjecture that this loss is due to scattering from and/or waveguide coupling into the beam.

Combining measurements of $g_0$ and $\kappa$ with typical room temperature mechanical damping rate of $\Gamma_\t{m}=2\pi\cdot 15$ Hz (we observed no change in $\Gamma_\t{m}$ for small beam-disk seperation, suggesting that squeeze-film gas damping \cite{verbridge2008size} was not a factor), the single-photon cooperativity is observed to approach $\mathcal{C}_0\sim1$. % for the samples tested in Fig. \ref{fig:char_lateral}.
This value is limited by the unfavorable scaling of $g_0^2/\kappa$ as $g_0$ begins to saturate.  Despite this limitation, the inferred $\mathcal{C}_0$ represents a nearly 50 dB increase over our prior chip-scale implementation \cite{gavartin_hybrid_2012}, owing to the combined 100-fold increase of $g_0$ and 10-fold reduction in $\kappa$.  Increase $g_0$ is due to the precise vertical and lateral positioning of the beam afforded by CMP and e-beam processing.  Reduced $\kappa$ is due to greater isolation of the disk during beam patterning, making use of the poly-Si sacrificial layer.  \fref{fig:char_lateral}b suggests that $\kappa$ is ultimately dominated by beam-induced scattering/absorption loss, rather than deterioration of intrinsic disk loss (\fref{fig:design_kappa}), implying that an additional 10-fold reduction in $\kappa$ may yet be realized with appropriate beam shaping/positioning.

%\vfill \pagebreak

\subsection{$g_0$ and $\mathcal{C}_0$ versus beam width and disk thickness}\label{sec:char_overlap}

Wider beams ($w\sim \lambda$) and thinner disks $(t_\t{d}<\lambda)$ were fabricated in an attempt to increase $g_0$ and $\mathcal{C}_0$ (see \eqref{eq:Gapprox}).  Measurements of $\{g_0,\mathcal{C}_0\}$ vs $w$ for two microdisk thicknesses, $t_\t{d}\approx0.43\;\t{and}\;0.63\;\mu\t{m}$, are shown in \fref{fig:char_lateral}. Fixed dimensions of the nanobeam and microdisk are $\{t,l\}\approx \{0.06,60\}\;\mu\t{m}$ and $\{r_\t{d},\theta\}\approx\{15\;\mu\t{m},30\;\t{deg.}\}$, respectively.  The lateral beam position was chosen to maximize $g_0$ for the $0.4\;\mu\t{m}$-wide beam (see \fref{fig:char_lateral}). For the TE optical modes studied,  a roughly 2$\times$ increase in $g_0$ was observed for the $30\%$ thinner disk.  In both cases, $g_0$ scaled roughly linearly for widths $w\in[0.4,1]\;\mu$m.  $\mathcal{C}_0$ also increased with $w$, roughly in proportion to $g_0^2$, for both $t_\t{d}$.  This is due to the fact that $\kappa$ (not shown) was roughly independent of $w$ for both disk thicknesses and a factor of four larger for the thinner disk.  The highest optomechanical coupling rate we have measured, $g_0\approx2\pi\cdot 150$ kHz, was for a 1 $\mu$m-wide beam coupled to a $0.43\;\mu\t{{m}}$-thick disk.  The highest cooperativities observed, $\mathcal{C}_0 > 2.5$, were for 1 $\mu$m-wide beams coupled to disks of both thicknesses.

%\vfill \pagebreak

\subsection{$g_0$ versus mechanical mode order}\label{sec:char_harmonics}

$g_0$ was also studied for higher order mechanical modes. As shown in \fref{fig:char_lateral}d
, $g_0$ decreases as the vibrational node spacing approaches the dimensions of the effective sampling length $l_\t{eff}$.  In this case the model in \secref{sec:g0model} --- which assumes rigid displacement of a beam with effective mass $m=\rho t w l/2$ --- breaks down.  A simple extension of the model is shown as a red line in \fref{fig:char_lateral}d. Here $m$ is computed with respect to optical-intensity-weighted displacement of the mechanical mode:  
\begin{subequations}\label{eq:g0vsn}
	\begin{align}
m &= \frac{\int_\t{beam}\rho |u(r)|^2 d^3 r}{\lvert\int_\t{beam} |E(r)|^2 u(r) d^3 r/\int_\t{beam} |E(r)|^2 d^3 r\rvert^2}\\ & \approx \frac{\rho t w l}{1-(-1)^n}\frac{1}{\t{sinc}^2\left(\frac{n\pi}{2}\frac{l_\t{eff}}{l}\right)}
\end{align}
\end{subequations}
where $\vec{u}(x,y,z) \approx\t{sin}(n\pi x/l)\hat{z}$ is the displacement profile of the $n^\t{th}$-order out-of-plane flexural mode. The latter expression is appropriate when the transverse dimensions of the beam are much smaller than that of the WGM, and assumes that the intensity distribution sampled by the beam is uniform along the beam axis with an effective sampling length $l_\t{eff}$.  Using $\Omega_\t{m}\propto n$ gives $g_0^{(n)}/g_0^{(0)}\approx \lvert\t{sinc}\left(\frac{n \pi}{2}\frac{l_\t{eff}}{l}\right)\rvert/\sqrt{n}$ for odd $n$ and 0 for even $n$.  The model shown in \fref{fig:char_lateral}d agrees quantitatively with experiment assuming an effective length of $l_\t{eff}=9.6\;\mu$m as the only free parameter.  A simple route to increasing $g_0$ would be to remove mass from the beam outside of the effective sampling length (see \fref{fig:outlook}).

\section{Displacement sensitivity}\label{sec:disp}

As an illustration of device performance, we use the microdisk to perform a cavity-enhanced interferometric measurement of the beam's displacement. For this purpose, the fiber taper and microdisk are embedded in one arm of a length- and power-balanced homodyne interferometer (\fref{fig:char_experimentalsetup}).  The cavity is driven on resonance using the Pound-Drever-Hall technique.  A piezoelectric mirror is use to stabilize the interferometer path length difference so that the homodyne photocurrent is proportional to the phase of the transmitted cavity field.

\begin{figure*}[th!]
	\centering
	\includegraphics[width=1.0\linewidth]{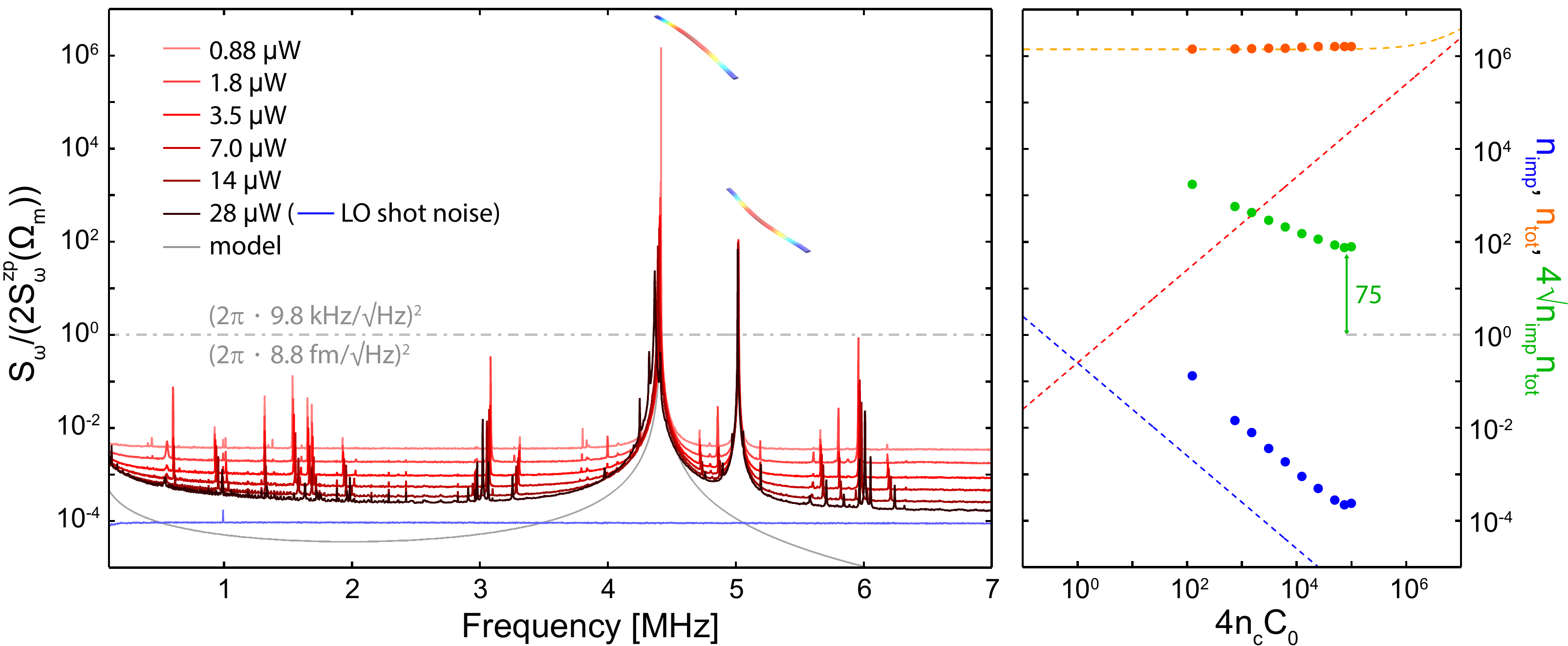}
	\caption{(a) Nanobeam displacement noise, measured by balanced homodyne detection of the microdisk output field, for various input powers.  Noise spectra are expressed in units relative to the cavity frequency noise produced by one phonon of fundamental out-of-plane vibration, $2S_\omega^{zp}(\Omega_\t{m})=(2\pi\cdot9.8\,\t{kHz}/\sqrt{\t{Hz}})^2$, where $\Omega_\t{m}=2\pi\cdot4.4$ MHz. For large power, the fundamental noise peak is shifted and broadened by optical spring softening and damping, respectively.  The peak at 5 MHz is due to thermal motion of the fundamental in-plane mode.  The gray curve is a model for the intrinsic thermal motion of the fundamental mode (\eqref{eq:Saulson}). (b) Measured phonon equivalent displacement, $n_\t{tot} = S_\omega(\Omega_\t{m})/2S_\omega(\Omega_\t{m})^\t{zp}$, displacement imprecision, $n_\t{imp}$, and their geometric mean versus intracavity photon number.  Dashed lines denote ideal values for $n_\t{tot}=n_\t{th}+n_\t{ba}+n_\t{imp}$ (orange), $n_\t{ba}=\mathcal{C}_0n_\t{c}$ (red), and $n_\t{imp}=1/16\mathcal{C}_0n_\t{c}$ (blue), using $n_\t{th}\approx1.4\cdot10^6$ and $\mathcal{C}_0=0.8$. Green arrow  indicates proximity to the uncertainty  limit, $4\sqrt{n_\t{imp}n_\t{tot}}\ge 1$.}\label{fig:disp}
	\vspace{-0pt}
\end{figure*}

Displacement noise spectra are shown in \fref{fig:disp} for a $\{l,w,t\}=\{60,0.4,0.06\}\;\mu$m beam with optomechanical parameters $\{\Omega_\t{m},\,\Gamma_\t{m},\,\kappa,\, g_0,\,\mathcal{C}_0\}\approx\{2\pi\cdot4.4\,\t{MHz},\,2\pi\cdot12\,\t{Hz},\,2\pi\cdot350\,\t{MHz},\,2\pi\cdot30\,\t{kHz},\,0.8\}$.  (Here $\kappa$ corresponds to the critically-coupled cavity linewidth and other parameters correspond to the fundamental out-of-plane mechanical mode.)  For the measurements shown, the cavity was critically coupled and the power of the input field was swept from $0.1-30\;\mu$W. The homodyne photocurrent noise spectrum is plotted in units relative to the signal produced by a phonon of displacement $2S_\omega^\t{zp}(\Omega_\t{m})\approx(2\pi\cdot9.8\,\t{kHz}/\sqrt{\t{Hz}})^2$ (equivalent to $2S_x^\t{zp}(\Omega_\t{m})\approx(2\pi\cdot8.8\,\t{fm}/\sqrt{\t{Hz}})^2$ assuming $x_\t{zp}=25$ fm).  In these units, the magnitude of the fundamental thermal noise peak (neglecting photothermal or dynamical back-action) is equal to the effective thermal occupation $n_\t{tot} = n_\t{th}+n_\t{ba}+n_\t{imp}$, where $n_\t{th} \approx k_\t{B}T/\hbar\Omega_\t{m}\approx 1.4\cdot 10^6$ is the ambient bath occupation, $n_\t{ba}$ is the effective thermal bath occupation associated with classical and quantum measurement back-action, and $n_\t{imp}\equiv S_\omega^\t{imp}(\Omega_\t{m})/2S_\omega^\t{zp}(\Omega_\t{m})$ is the apparent thermal occupation associated with the measurement imprecision, $S_\omega^\t{imp}$.  The noise spectra are calibrated by bootstrapping a low power measurement to $S_\omega(\Omega_\t{m})/(2S_\omega^\t{zp}(\Omega_\t{m})\approx 2 n_\t{th}$. For larger optical powers, dynamic spring/damping forces modify the peak. At the highest optical powers, the displacement imprecision in the vicinity of $\Omega_\t{m}$ is estimated (from the saddle at 2.5 MHz) to be $n_\t{imp}\approx 2.5\cdot10^{-4}$, while the shot-noise equivalent displacement is  $n^\t{(shot)}_\t{imp}\approx 1.0\cdot10^{-4}$.  These correspond to imprecisions 30 and 34 dB below that at the SQL ($n_\t{imp}=0.25$), respectively.  The absolute magnitude of the extraneous imprecision, $2S^\t{zp}_\omega(\Omega_\t{m})\cdot(n_\t{imp}-n_\t{imp}^\t{(shot)})\approx (2\pi\cdot 120\,\t{Hz}/\sqrt{\t{Hz}})^2$, is consistent with a mixture of ECDL noise ($\sim30\,\t{Hz}/\sqrt{\t{Hz}}$ \cite{wilson2015measurement}), TRN ($\sim10\,\t{Hz}/\sqrt{\t{Hz}}$ \cite{anetsberger_near-field_2009}), and off-resonant thermal noise ($\sim70\,\t{Hz}/\sqrt{\t{Hz}}$). The latter is estimated using the `structural damping' model of Saulson \cite{saulson1990thermal},
\begin{equation}\label{eq:Saulson}
\frac{S_\omega(\Omega)}{2S^\t{zp}_\omega(\Omega_\t{m})} \approx n_\t{th}\frac{\Omega_\t{m}}{\Omega}\frac{\Gamma_\t{m}^2\Omega_\t{m}^2}{(\Omega^2-\Omega_\t{m}^2)^2+\Gamma_\t{m}^2\Omega_\t{m}^2}\lesssim \frac{7n_\t{th}}{Q_\t{m}^2},
\end{equation}
shown in gray in \fref{fig:disp}, for $Q_\t{m} = \Omega_\t{m}/\Gamma_\t{m} = 3.7\cdot 10^5$.  

The total efficiency of the measurement is estimated by comparing the power dependence of the imprecision ($n_\t{imp}$), the effective thermal bath occupation ($n_\t{tot}$), and their geometric mean $\sqrt{n_\t{imp}n_\t{tot}}$ to the ideal values $1/(16\mathcal{C}_0 n_\t{c})$, $\mathcal{C}_0 n_\t{c}$, and $1/4$, respectively, where the last case represents the Heisenberg uncertainty limit \cite{wilson2015measurement}. As shown on the right hand side of \fref{fig:disp}, the imprecision is a factor of 40 larger than ideal, due to a combination of cavity loss ($50\%$), taper loss ($\sim10\%$), homodyne detector loss/misalignment, and optical mode splitting \cite{wilson2015measurement}.  The effective thermal bath occupation is inferred by fitting to the off-resonant tail of the fundamental noise peak (to avoid the systematic error due to optical damping).  From these fits we infer a heating of $C_\t{0}^\t{ext}\equiv (n_\t{tot}-n_\t{th})/n_\t{c}=1.6$, two times larger than expected due to quantum measurement back-action. The imprecision-back-action product is constrained, at high powers, to $4\sqrt{n_\t{imp}n_\t{tot}}\approx75$, due to the saturation of the measurement imprecision.  To the best of our knowledge, this represents the closest approach to the uncertainty limit for a room temperature mechanical oscillator.

\section{Summary and outlook}\label{sec:conclusion}
 
We have presented a method to heterogeneously integrate a high-stress, Si$_3$N$_4$ nanobeam within the evanescent near-field of a SiO$_2$ microdisk. Building on earlier strategies \cite{anetsberger_near-field_2009,gavartin_hybrid_2012}, the principle advance is a novel vertical integration technique which preserves the high $Q/$(mode volume) ratio of each resonator while enabling the beam and the disk to be separated by as little as 25 nm --- significantly smaller than the evanescent decay length of the microdisk's WGMs.  Samples of various dimensions were fabricated and characterized.  Simultaneously low mechanical loss, $\Gamma_\t{m}=2\pi\cdot(10-100)\;\t{Hz}$, low optical loss, $\kappa=2\pi\cdot(100-1000)\;\t{MHz}$, and large optomechanical coupling rates, $g_0 = 2\pi\cdot (10-100)\;\t{kHz}$, were measured, corresponding to room temperature single-photon cooperativities as high as $\mathcal{C}_0 \equiv 4g_0^2/{\Gamma_\t{m}\kappa} = 2$.  

The reported system holds particular promise as a quantum-limited displacement sensor, owing to the large vacuum displacement of the nanobeam 
and the high power handling capacity of the microdisk.  For a typical device, possessing $\{\Omega_\t{m},\Gamma_\t{m},\kappa_0, g_0\}\approx 2\pi\cdot\{4.5\;\t{MHz},15\;\t{Hz}, 500\;\t{MHz},50\;\t{kHz}\}$, the resonant vacuum displacement noise, $S_\omega^\t{zp}(\Omega_\t{m}) = 4g_0^2/\Gamma_\t{m} \approx (2\pi\cdot 26\;\t{kHz}/\sqrt{\t{Hz}})^2$, is orders of magnitude larger than major sources of imprecision --- such as laser frequency and thermorefractive noise  \cite{anetsberger_measuring_2010} --- and commensurate with shot noise for an ultra-low intracavity photon number of $n_\t{c}=1/(16 \mathcal{C}_0)= 0.05$ \cite{wilson2015measurement}.  Operating a similar device at 4 K with $n_\t{c}\sim 10^5$ (corresponding to $P\sim 100\;\mu\t{W}$ when critically coupled to the fiber waveguide), we were able to achieve a displacement imprecision $n_\t{th}$ times (43 dB) below $S_\omega^\t{zp}$, while maintaining an imprecision-back-action product within a factor of 5 of the uncertainty limit \cite{wilson2015measurement}. This regime of `efficient' measurement --- characterized by the ability to resolve a phonon-equivalent displacement in the thermal decoherence time --- enabled us to feedback cool the mechanical mode to near its ground state \cite{wilson2015measurement}, and might be extended to other quantum control tasks, such as squeezed-state preparation \cite{szorkovszky2011mechanical}.

\begin{figure}[tr!]
	\centering
	\includegraphics[width=1\linewidth]{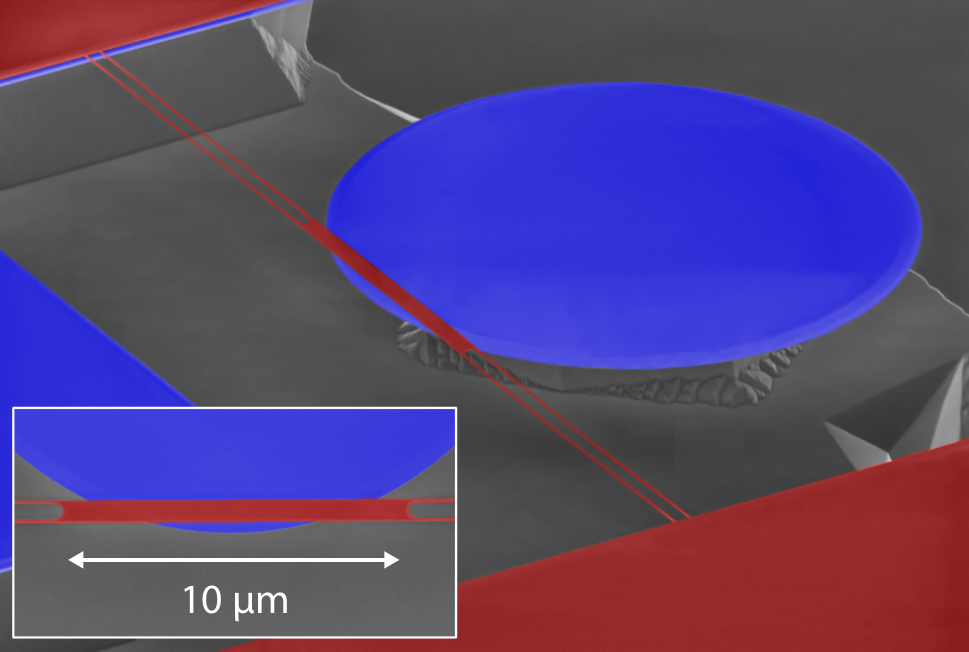}
	\caption{Variation on a theme: suspending the nanobeam from tethers enables higer $g_0$ by reducing mass without changing optomechanical mode overlap.  Here the central beam coincides with the effective sampling length of the optical mode.}
	\label{fig:outlook}
	\vspace{-10pt}
\end{figure}

An intriguing question is whether the reported device may be used to realize Heisenberg-limited displacement measurements at room temperature.  For the radio frequency oscillators under study ($n_\t{th}\sim10^6$), the main challenges are (1) pumping the cavity with $n_\t{c}=n_\t{th}/C_{0}\sim 10^6$ photons in order to achieve the necessary measurement strength (characterized by a phonon-equivalent radiation pressure shot noise of $n_\t{ba}=\mathcal{C}_0 n_\t{c}>n_\t{th}$), (2) reducing extraneous sources of measurement imprecision to $S_\omega^\t{zp}/n_\t{th} < (2\pi\cdot 10\;\t{Hz}/\sqrt{\t{Hz}})^2$, and (3) reducing extraneous heating to below $n_\t{ba}$.  Because of the (blue-stable) thermal self-locking effect in room temperature SiO$_2$ microcavities \cite{arcizet2009cryogenic}, the first requirement %(corresponding to $P\sim 1$  mW for critically coupling with $\kappa\sim 1$ GHz) 
is expected to be limited by parametric radiation pressure instabilities, requiring active feedback damping.  (Taking a different approach, cross-correlation techniques may be employed to detect radiation pressure shot noise at the 1$\%$ level \cite{verlot2009scheme}, significantly relaxing associated demands on input power and active stabilization.)  For microdisks with dimensions studied here, the second requirement is expected to be limited by thermorefractive noise at the level of $S^\t{trn}_\omega\sim (2\pi\cdot10\;\t{Hz}/\sqrt{\t{Hz}})^2$ \cite{anetsberger_measuring_2010}, an impressive 60 dB lower than $S_\omega^\t{zp}$.  Reaching $S^\t{trn}_\omega<S_\omega^\t{zp}/n_\t{th}$ would require a moderate increase in $g_0^2\cdot Q_\t{m}$ (for instance, by using lower-mass, ``tethered" beams \cite{krause2015optical}; see \fref{fig:outlook}).  The third requirement depends on the details of the nanoscale heat transfer process.  At 4 K, we have observed photothermal heating consistent with an extraneous cooperativity of $\mathcal{C}_0^\t{ext}\equiv n_\t{ba}^\t{ext}/n_\t{c} \sim 1$ \cite{wilson2015measurement}; we anticipate this heating to reduce to tenable levels ($\mathcal{C}_0^\t{ext}<\mathcal{C}_0$) at room temperature, provided that the underlying process is related to the temperature-dependent thermal conductivity of amorphous glass \cite{pohl2002low}.  Preliminary room temperature measurements, discussed in \secref{sec:disp}, suggest that $\mathcal{C}_0^\t{ext}\sim\mathcal{C}_0$ can be met for a moderate $\mathcal{C}_0\sim 0.8$. 

In addition to high cooperativity, the evanescent sensing platform and reported fabrication method have as a compelling feature the ability to incorporate new materials and/or planar geometries above a high-Q microdisk with nanometric precision.  This capability opens the door to a variety of ``hybrid" sensing applications.  For example, the system may be electrically functionalized by vertically integrating  the beam into a parallel plate capacitor, or into the gradient field between two closely spaced electrodes \cite{unterreithmeier2009universal,okamoto2015strongly}.  This interface --- which need not compromise the mechanical quality of the beam \cite{yu2012control} --- can form the basic building block of a high-efficiency electro-optic converter, with applications such as precision radio wave sensing \cite{bagci2014optical}.  Pushed to a different extreme, two dimensional materials such as graphene or MoS$_2$ may be integrated with a microdisk by using the SiN film as a sacrificial substrate \cite{schmid2014single}.  Another intriguing possibility, in conjunction with the preparation of low entropy mechanical states using measurement-based feedback, is to functionalize the beam with a two-level system, such as an NV center embedded in a diamond nanocrystal \cite{arcizet2011single}. Nanobeams integrated with microdisks may also serve as a platform for remotely-coupled ``atom-optomechanics" \cite{hammerer2009establishing,jockel2014sympathetic}, taking advantage of the low oscillator mass, high cavity finesse, and recent developments in fiber-based atom traps \cite{vetsch2010optical}.   Finally, we note that the ability to perform broadband, thermal-noise-limited measurements of high-$Q$ nanomechanical oscillators may help shed light on the microscopic origin of intrinsic damping \cite{groblacher_observation_2015}.
\\
\begin{center}
\textbf{ACKNOWLEDGMENTS}
\end{center}

We acknowledge fabrication advice from E. Gavartin in early stages of the project. All samples were fabricated at the CMi (Center of MicroNanotechnology) at EPFL. Research was funded by an ERC Advanced Grant (QuREM), by the DARPA/MTO ORCHID program, the Marie Curie Initial Training Network `Cavity Quantum Optomechanics' (cQOM), the Swiss National Science Foundation and through support from the NCCR of Quantum Engineering (QSIT). D.J.W. acknowledges support from the European Commission through a Marie Skłodowska-Curie Fellowship (IIF project 331985).
	%\item[Contributions] R.S., A.G. and D.J.W. contributed to design and initial characterization of the device.  R.S. developed, implemented, and optimized the fabrication process.  A.G. developed and implemented a numerical model to aid in device optimization.  D.J.W, V.S., and N.P. conceived of, designed, and performed the experiment.  V.S. and D.J.W. analysed the data with support from N.P.. V.S. developed the theoretical framework of the experiment with support from D.J.W. and N.P..  V.S. wrote Sec. I-II of the supplementary information.  R.S., V.S., D.J.W., and A.G. wrote Sec. III of the supplementary information.  D.J.W, N.P., and V.S wrote Sec. IV-V of the supplementary information.  D.J.W. wrote the main text with support from all other authors.   T.J.K. oversaw all aspects of the work.
	%\item[Competing Interests] The authors declare that they have no competing financial interests.
	%\item[Correspondence] Correspondence and requests for materials should be addressed to T. J. Kippenberg ~(email: tobias.kippenberg@epfl.ch).

}
%\vfill \pagebreak
\bibliographystyle{apsrev4-1}

%\bibliography{ref1}

\end{document}